%% file: paper_v5_2.tex
\documentclass[a4paper,12pt]{article}
\usepackage{a4}

\usepackage{amsmath}
\usepackage{amsfonts}
\usepackage{amssymb}
\usepackage{graphicx}
\usepackage{array}

\usepackage{color}
\newcommand{\FPa}{\input{fixpts_10}}
\newcommand{\FPas}{\input{fixpts_11}}

\newcommand{\FPnone}{\input{fixpts_none}}
\newcommand{\FPeq}{\makebox(30,64){$= 0$}}
\newcommand{\FPneq}{\makebox(30,64){$\neq 0$}}

\newcommand{\M}{\mathcal{M}}
\newcommand{\N}{\mathcal{N}}
\newcommand{\R}{\mathcal{R}}
\newcommand{\T}{\mathbb{T}}
\newcommand{\Z}{\mathbb{Z}}
\newcommand{\Rs}{\Box\hspace{-0.65mm}\Box}
\newcommand{\Ra}[1]{\begin{array}{l}\Box\\[-2.27mm]\Box_{#1}\end{array}}
\newcommand{\Ratext}{\begin{array}{l}\Box\\[-1.44mm]\Box\end{array}}

\allowdisplaybreaks[2]
\numberwithin{equation}{section}

\begin{document}


\vspace*{-1.5cm}
\begin{flushright}
  {\small
  MPP-2006-38 \\
  LMU-ASC 24/06 \\
  hep-th/0604033 \\
  }
\end{flushright}

\vspace{1.5cm}
\begin{center}
  {\LARGE
  Intersecting D-Branes on Shift $\Z_2\times\Z_2$ Orientifolds \\
  }
\end{center}

\vspace{0.25cm}
\begin{center}
  {\small
  Ralph Blumenhagen$^1$ and Erik Plauschinn$^{1,2}$ \\
  }
\end{center}

\vspace{0.1cm}
\begin{center}
  \emph{$^{1}$
  Max-Planck-Institut f\"ur Physik, F\"ohringer Ring 6, \\
  80805 M\"unchen, Germany } \\
  \vspace{0.25cm}
  \emph{$^{2}$ Arnold-Sommerfeld-Center for Theoretical Physics, \\
  Department f\"ur Physik, Ludwig-Maximilians-Universit\"at  M\"unchen, \\
  Theresienstra\ss e 37, 80333 M\"unchen, Germany} \\
\end{center}

\vspace{-0.1cm}
\begin{center}
  \tt{
  blumenha,plausch@mppmu.mpg.de \\
  }
\end{center}

\vspace{1.5cm}
\begin{abstract}
\noindent We investigate $\Z_2\times\Z_2$ orientifolds with group actions
involving shifts. A complete classification of possible geometries is presented where also previous work by other authors is included in a unified framework from an intersecting D-brane perspective. In particular, we show that the additional shifts not only determine the topology of the orbifold but also independently the presence of orientifold planes.

\noindent In the second part, we work out in detail a basis of homological three cycles on shift $\Z_2\times\Z_2$ orientifolds and construct all possible fractional D-branes including rigid ones. A Pati-Salam type model with no open-string moduli in the visible sector is presented.
\end{abstract}

\thispagestyle{empty}
\clearpage
\tableofcontents


\section{Introduction}

Both heterotic and Type I string compactifications show
the salient features of the Standard Model of particle physics as
the emergence of gauge symmetry, the existence of three generations
of chiral matter or the existence of vector-like Higgs particles.
In each class one can construct concrete models which satisfy
part of the constraints one imposes from the low energy
physics we know up to the energy scale of $10^2$ GeV. However, all
concrete models studied so far fail at a certain step if
more refined Standard Model features  are required.

There are both indications from particle physics and from string theory
that supersymmetry should play an important role for the physics
beyond the Standard Model. For this reason, in all consistent  string
theory
examples one  first
focuses on the construction of supersymmetric string models and
then breaks supersymmetry at low energies in a controllable way.
While the $E_8\times E_8$ heterotic string appears to be more
natural for the embedding of $SU(5)$ and $SO(10)$ GUT models,
the Type I string (see \cite{Angelantonj:2002ct} for a review)
  with intersecting D-branes might be considered more
natural for directly providing the Standard Model or a Pati-Salam
type gauge symmetry (see e.g. \cite{AU03,Lust:2004ks,Blumenhagen:2005mu}
for reviews).
In particular the appearance of $U(1)$ gauge
symmetries is almost inevitable in D-brane models.
As long as we do not know what the gauge group at the string scale
really is, both approaches should be studied to see how close
these vacua can resemble the Standard Model.

Supersymmetric intersecting D-brane models have mostly been studied
on toroidal   orbifolds
\cite{bgk99a,Forste:2000hx,Cvetic:2001tj,Cvetic:2001nr,bgo02,Cvetic:2002pj,GH03,cp03a,Larosa:2003mz,Cvetic:2004ui,Honecker:2004kb,Blumenhagen:2004di,Cvetic:2004nk,Dudas:2005jx,Blumenhagen:2005tn,Chen:2005ab}.
Following the approach of \cite{Blumenhagen:2002wn}, some of the basic
properties of
such models are determined just by the topology of the underlying brane
configurations. For instance the chiral massless spectrum can be computed
once
one knows
the topological intersection form on  $H^3(X,\Z )$ with $X$ denoting  the
orbifold space.
The fixed loci of the orbifold action can be appropriately taken care of
by
introducing twisted 3-cycles.
  In addition one needs the action of the orientifold on the homological
basis.

In a completely satisfactory model one has to implement  a mechanism
to freeze the extra massless fields (moduli), which are ubiquitously
present in all string compactifications. Here one is confronted with both
closed and open string moduli, where the latter parameterise the
deformations of the D-branes in addition to possible continuous
Wilson lines along the D-brane. For unitary gauge group on the D-brane,
these scalars transform in the adjoint representation of the gauge group,
so that, being light, they are contributing to the one-loop beta function
for the gauge couplings\footnote{For a discussion of gauge coupling unification for intersecting D-brane models see \cite{bls03}.}.
The main insight into this problem during
the last years has been that turning on fluxes generates scalar potentials
which can freeze part or even all of the above moduli.
However, one eventually has to decide whether the mass scale of
for instance the open string moduli is sufficiently large to not destroy
asymptotic freedom of $SU(3)_c$.

From this perspective it is also interesting to have D-branes which
do not have any deformations in the first place.
For space-time filling
intersecting D6-branes this means that they wrap so-called rigid 3-cycles
in the internal manifold.
These brane moduli
are counted by the number of closed one-cycles on the three cycle
$\Gamma$ the D6-brane is wrapping, i.e. by the Betti number $b^1(\Gamma)$.
In most orbifold examples studied so far such
rigid 3-cycles are absent.
Consider for instance the mostly studied  $\Z_2\times \Z_2$ orbifold.
Type IIA D-brane models on this orbifold with Hodge numbers
$(h^{2,1},h^{1,1})=(3,51)$ have been first discussed in
\cite{Forste:2000hx,Cvetic:2001tj,Cvetic:2001nr}. Here there are only
eight 3-cycles the D6-branes can wrap around leading to only four
Ramond-Ramond
(R-R) tadpole constraints in addition to the same number of torsion
K-theory
constraints \cite{MarchesanoBuznego:2003hp}. Therefore, a lot of
supersymmetric D-brane models exist, whose statistical distributions have
been
discussed in \cite{Blumenhagen:2004xx}. However the D6-branes do not
wrap rigid 3-cycles so that one always gets massless adjoint matter in
these
cases.
On the contrary, it has been shown in \cite{Dudas:2005jx,Blumenhagen:2005tn} that
on the other $\Z_2\times \Z_2$ orbifold with Hodge numbers
$(h^{2,1},h^{1,1})=(51,3)$ there exist a large number of rigid 3-cycles.
However, there is a certain price to pay, namely that some of the
orientifold
planes change sign and that one gets 52 tadpole conditions, which makes it
much
harder to find semi-realistic models
\cite{Blumenhagen:2005tn,liu_cvetic_privcom}.

Sort of in the middle between these two extreme models are shift
$\Z_2\times
\Z_2$ orbifolds with Hodge numbers $(h^{2,1},h^{1,1})\in\{
(3,3),(11,11),(19,19)\}$. Therefore, one might hope to still get only a
moderate number of tadpole conditions with the possibility of having rigid
cycles. The aim of this paper is to revisit  these models from this
perspective and thereby systemise and generalise the earlier results
\cite{Antoniadis:1998ep,Antoniadis:1999ux,Pradisi:2002vu,Larosa:2003mz,Pradisi:2003ct}.
The  $\Z_2\times\Z_2$ orbifold is one of the standard examples for
orbifold
compactification in string theory. The usual realization of the group
action is
a combination of rotations by $\pi$ of the internal coordinates. However,
this
is not the most general choice because the rotations may be accompanied by
translations. The first type of such shifts would be $z\to z+\delta$ which
has
already been studied in
\cite{Antoniadis:1998ep,Antoniadis:1999ux,Pradisi:2002vu,Larosa:2003mz,Pradisi:2003ct,Angelantonj:2005hs}.
But there is a second type of shifts $z\to -z+\kappa$ which does not
change the topology of the orbifold but has implications at the level of
the
orientifold. The third possibility for translations are shifts
accompanying the
orientifold projection $\Omega\R$. In this paper we systematically
analyse the consequences of
the various shifts on the geometry, in particular we show that they modify
the
presence of orientifold planes in the various sectors independent of the
orbifold topology.

This paper is organised as follows. In section \ref{sec_2} we analyse and
classify the possible shift $\Z_2\times\Z_2$ orientifold geometries. In
section
\ref{sec_3}, we work out the possible fractional D6-branes and the
formulas
for the generic chiral massless spectrum. In addition, we derive the R-R
tadpole cancellation conditions including their explicit dependence on the
various shifts. Finally, in section \ref{sec_4} we present a Pati-Salam 
type
model with no open string-moduli in the visible sector for one of the 
shift
orientifold backgrounds. Section \ref{sec_5} contains a summary and 
outlook.


\section{Classification of Shift $Z_2\times Z_2$ Orientifolds}
\label{sec_2}

In this section, the possible shift $\Z_2\times\Z_2$ orbifold and orientifold geometries are characterized. To make the setup concrete, the $10$-dimensional space-time is assumed as $\mathbb{R}^{1,3}\times\M_6$ and the $6$-dimensional, compact, internal space $\M_6$ is chosen as the orientifold
\begin{equation}\label{internal_space}
  \M_6=\frac{\T^2\times\T^2\times\T^2}
  {\Z_2\times\Z_2+\Omega\R\cdot\Z_2\times\Z_2}
\end{equation}
where the action of $\Z_2\times\Z_2$ and $\R$ will be specified in the following. The underlying string theory is type IIA.


\subsection{Orbifold Background}
\label{sec_2_orbifold_background}

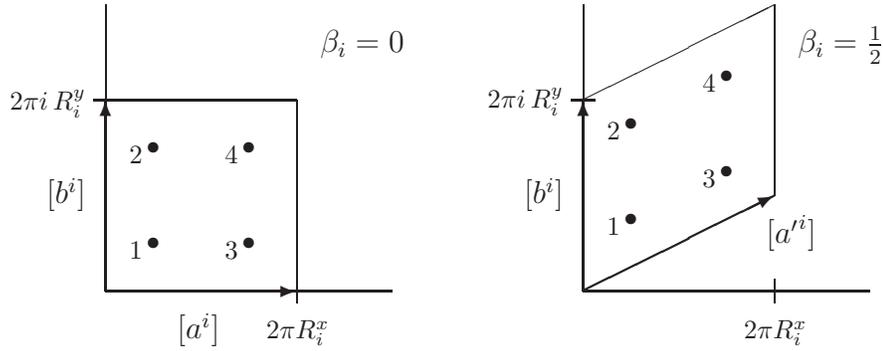
\begin{figure}[ht]\begin{center}
\input{figure_01}
\caption{Choices of complex structures, fixed points and fundamental cycles on
one $\T^2$ factor.\label{fig_torus}}
\end{center}\end{figure}

Let us consider first the factorizable six-torus $\T^6=\T^2\times\T^2\times\T^2$. On each of the $\T^2$ factors one can introduce complex coordinates $z_i$ where $i=1,2,3$. Our choice of complex structures is specified by identifying $z_i\cong z_i+n\,e^x_i+m\,e^y_i$ where $n,m\in\Z$ and the $e^a_i$ are defined as
\begin{equation}\label{torus_basis}
  e^x_i=2\pi\,R_i^x+i\,\beta_i\,2\pi\,R_i^y \qquad\mbox{and}\qquad
  e^y_i=i\,2\pi\,R_i^y \,.
\end{equation}
In general, the basis $\{e^x_i,e^y_i\}$ of the factorizable $\T^6$ for the $\Z_2\times\Z_2$ orbifold is arbitrary. But later an anti-holomorphic involution $\R$ will be introduced which is compatible with \eqref{torus_basis} only for $\beta_i=0,\,1/2$. These definitions are illustrated in figure \ref{fig_torus}.

\bigskip
Next, consider the orbifold $\T^6/\Z_2\times\Z_2$ where the six-torus is again
factorizable. The group $\Z_2\times\Z_2$ acts on the coordinates $z_i$ in each
two-torus and since we are interested in a group action involving shifts, we make the following general choice for the $\Z_2$ generators $\Theta$ and $\Theta'$
\begin{equation}\begin{array}{c}
  \Theta :\left\{ \begin{array}
    {r@{\hspace{1pt}}r@{\hspace{1pt}}l@{\hspace{1pt}}l}
    z_1 \to&-z_1&         &+\kappa_1\\
    z_2 \to&-z_2&         &+\kappa_2\\
    z_3 \to& z_3&+\delta_3&
  \end{array} \right. , \qquad
  \Theta':\left\{ \begin{array}
    {r@{\hspace{1pt}}r@{\hspace{1pt}}l@{\hspace{1pt}}l}
    z_1 \to& z_1&+\delta_1           \\
    z_2 \to&-z_2&+\delta_2&+\kappa_2 \\
    z_3 \to&-z_3&         &+\kappa_3
  \end{array} \right. , \vspace{15pt} \\
  \Theta\Theta':\left\{\begin{array}
    {r@{\hspace{1pt}}r@{\hspace{1pt}}l@{\hspace{1pt}}l}
    z_1 \to&-z_1&+\delta_1&+\kappa_1\\
    z_2 \to& z_2&+\delta_2          \\
    z_3 \to&-z_3&+\delta_3&+\kappa_3
  \end{array} \right. \,.
\end{array}
\label{orb_group}
\end{equation}
The group element $\Theta\Theta'$ is included for completeness and the shifts are defined as $\delta_i=\delta^x_i\,e^x_i+\delta^y_i\,e^y_i$ and $\kappa_i=\kappa^x_i\,e^x_i+\kappa^y_i\,e^y_i$. For consistency, one has to require that all group elements square to the identity modulo the identifications on the torus. This implies that the $\delta_i$ are restricted and in summary one finds
\begin{equation}
  \delta^a_i=0,1/2 \qquad \mbox{and} \qquad \kappa^a_i\in [0,1)
\end{equation}
where $a=x,y$. It is important to realize that the $\delta_i$ determine different orbifold configurations and that at the orbifold level all choices of $\kappa_i$ lead to equivalent geometries. However, as it will be shown in section \ref{sec_2_orientifold_background}, at the level of orientifolds also the $\kappa_i$ determine different configurations.

\bigskip
Depending on the shifts $\delta_i$, the orbifold group \eqref{orb_group} can have fixed points. It is easy to see that for instance the action $\mathbb{I}\,z=-z+\kappa$ on one $\T^2$ leaves the four points
\begin{center}\begin{tabular}{lcl}
  $1/2\,\kappa$,&& $1/2\,\kappa+1/2\,e^x$, \\
  $1/2\,\kappa+1/2\,e^y$ & and & $1/2\,\kappa+1/2\,(e^x+e^y)$
\end{tabular}\end{center}
invariant. On the other hand, the action $\mathbb{S}\,z=z+\delta$ with $\delta\neq 0$ has no fixed points. For the factorizable $\T^6$ this implies that if $\delta_3=0$ then four points in $\T^2_1$, four points in $\T^2_2$ and the whole $\T^2_3$ are invariant under $\Theta$. Therefore, in this case, $\Theta$ leaves 16 $\T^2$ invariant. In summary, one finds that
\begin{center}\begin{tabular}{l}
  if $\delta_3=0$ then $\Theta$ leaves 16 $\T^2_3$ invariant, \\
  if $\delta_1=0$ then $\Theta'$ leaves 16 $\T^2_1$ invariant and \\
  if $\delta_2=0$ then $\Theta\Theta'$ leaves 16 $\T^2_2$ invariant.
\end{tabular}\end{center}
\begin{table}[p]
\begin{center}\fbox{
\begin{tabular}{l|cccccc}
  & $\delta_1$ & $\delta_2$ & $\delta_3$ &$\T^2_1$&$\T^2_2$&$\T^2_3$\\ \hline
  \makebox(15,64){0}&\FPneq&\FPneq&\FPneq&\input{fixpts_none}&\input{fixpts_none}&\input{fixpts_none}  \\ \hline
  \makebox(15,64){1}&\FPneq&\FPneq&\FPeq &\input{fixpts_10}   &\input{fixpts_10}   &\input{fixpts_none}  \\
                    &\FPneq&\FPeq &\FPneq&\input{fixpts_31}  &\input{fixpts_none}&\input{fixpts_31}    \\
                    &\FPeq &\FPneq&\FPneq&\input{fixpts_none}&\input{fixpts_21}  &\input{fixpts_20}     \\ \hline
                    &\FPneq&\FPeq &\FPeq &\input{fixpts_11_30} &\input{fixpts_10}   &\input{fixpts_30}     \\
  \makebox(15,64){2}&\FPeq &\FPneq&\FPeq &\input{fixpts_10}   &\input{fixpts_10_21} &\input{fixpts_20}     \\
                    &\FPeq &\FPeq &\FPneq&\input{fixpts_30}   &\input{fixpts_20}   &\input{fixpts_20_31}   \\ \hline
  \makebox(15,64){3}&\FPeq &\FPeq &\FPeq &\input{fixpts_10_30}  &\input{fixpts_10_20}  &\input{fixpts_20_30}
\end{tabular}}
\end{center}
\caption{Possible fixed point configurations. A dot indicates fixed points in the $\Theta$ sector, a cross in the $\Theta'$ sector and a square in the $\Theta\Theta'$ sector.} \label{tab_allFP}
\end{table}
In table \ref{tab_allFP} all combinations of the shifts $\delta_i$ and corresponding fixed points are listed. Since at the orbifold level the $\kappa_i$ lead to equivalent geometries, they have been set to zero. Also for simplicity, shifts with $\delta_i\neq0$ are represented by the diagonal shift and only tori with $\beta_i=0$ are shown. One can see that up to permutations of the $\T^2_i$, there are four distinct cases.
\begin{itemize}
\itemsep=0pt\parskip=0pt
\item[{\bf 0 :}] There are no fixed points. This case will not be considered.
\item[{\bf 1 :}] There are fixed points from one twisted sector.\footnotemark
                 \addtocounter{footnote}{-1}
\item[{\bf 2 :}] There are fixed points from two twisted sectors.\footnotemark
\item[{\bf 3 :}] There are fixed points from all three twisted sectors and no
                 shifts $\delta_i$. For $\kappa_i=0$, this case was already
                 analyzed from an intersecting brane perspective in
                 \cite{Forste:2000hx,Cvetic:2001tj,Cvetic:2001nr} for
                 the case without discrete torsion, and in
                 \cite{Blumenhagen:2005tn} for the
                 case with discrete torsion.
\end{itemize}
\footnotetext{Note that case 1 includes the $p_{23}$ model and case 2 includes the $p_3$ model of \cite{Antoniadis:1999ux,Larosa:2003mz}.}


\subsection{Hodge Numbers}

\begin{figure}[ht]
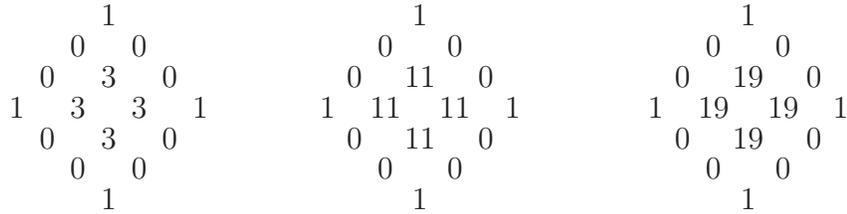
\begin{center}
  \renewcommand{\arraystretch}{0.8}
  \begin{tabular}{*{7}{c@{\extracolsep{2mm}}}}
    &&& 1 &&& \\
    && 0 && 0 && \\
    & 0 && 3 && 0 & \\
    1 && 3 && 3 && 1 \\
    & 0 && 3 && 0 & \\
    && 0 && 0 && \\
    &&& 1 &&&
  \end{tabular}
  \hspace{1cm}
  \begin{tabular}{c@{\extracolsep{1.5mm}}c@{\extracolsep{1.0mm}}
                   c@{\extracolsep{0.5mm}}c@{\extracolsep{0.5mm}}
                   c@{\extracolsep{1.0mm}}c@{\extracolsep{1.5mm}}c}
    &&& 1 &&& \\
    && 0 && 0 && \\
    & 0 && 11 && 0 & \\
    1 && 11 && 11 && 1 \\
    & 0 && 11 && 0 & \\
    && 0 && 0 && \\
    &&& 1 &&&
  \end{tabular}
  \hspace{1cm}
  \begin{tabular}{c@{\extracolsep{1.5mm}}c@{\extracolsep{1.0mm}}
                   c@{\extracolsep{0.5mm}}c@{\extracolsep{0.5mm}}
                   c@{\extracolsep{1.0mm}}c@{\extracolsep{1.5mm}}c}
    &&& 1 &&& \\
    && 0 && 0 && \\
    & 0 && 19 && 0 & \\
    1 && 19 && 19 && 1 \\
    & 0 && 19 && 0 & \\
    && 0 && 0 && \\
    &&& 1 &&&
  \end{tabular}
  \renewcommand{\arraystretch}{1}
\end{center}
\caption{Hodge diamond for zero, one and two twisted sectors with fixed points.} \label{fig_hodge_diamond}
\end{figure}

The topology and therefore the number of homological cycles on the different shift $\Z_2\times\Z_2$ orbifold configurations can be classified by the nontrivial Hodge numbers $(h^{2,1},h^{1,1})$. The contribution from the untwisted part is easily found as $(h^{2,1},h^{1,1})_{untw}=(3,3)$ and the resulting Hodge diamond is shown at the left in figure \ref{fig_hodge_diamond}.

The contribution from the twisted part can be understood as follows. Consider for instance the $\Theta$ sector with four fixed points in the first and four fixed points in the second two-torus. These 16 singularities in $\T^2_1\times\T^2_2$ can be resolved by a blow-up where one replaces each of the fixed points by a complex projective $1$-space $\mathbb{P}^1$. In this limit one obtains $\mathbb{K}3\times\T^2_3$. However, since $\mathbb{P}^1$ is isomorphic to the two-sphere $S^2$, there are now 16 additional $2$-cycles and also 16 additional $(1,1)$-forms. After taking the limit $\mathbb{K}3\to\T^2_1\times\T^2_2/\Z_2$ back to the orbifold, one is left with 16 collapsed $2$-cycles denoted by $e^{\Theta}_{ij}$ and with 16 localized $(1,1)$-forms denoted by $\omega^{\Theta}_{ij}$ where $i,j=1\ldots4$ label the fixed points as in figure \ref{fig_torus}. But, because of the action of $\Theta'$, for $\delta_1\neq0$ or $\delta_2\neq0$ there are only eight collapsed $(1,1)$-forms and eight collapsed $(2,1)$-forms which are invariant under the whole orbifold group $\Z_2\times\Z_2$. The invariant $(1,1)$- and $(2,1)$-forms coming from the $\Theta$ sector are
\begin{equation}\label{hodge_invariant_forms}
  \omega^{\Theta}_{ij}+\Theta'\left(\omega^{\Theta}_{ij}\right)
  \qquad \mbox{and} \qquad \left(\omega^{\Theta}_{ij}
  -\Theta'\left(\omega^{\Theta}_{ij}\right)\right)\wedge dz_3 \,.
\end{equation}
From \eqref{hodge_invariant_forms} one can conclude that, as long as the orbifold group \eqref{orb_group} contains at least one nontrivial shift $\delta_i$, each twisted sector with fixed points contributes $(h^{2,1},h^{1,1})_{tw}=(8,8)$ to the Hodge numbers. The resulting Hodge diamonds for zero, one and two twisted sectors with fixed points are displayed in figure \ref{fig_hodge_diamond}.

In the following, the above mentioned case 1 of table \ref{tab_allFP} with fixed points in one twisted sector will be labeled as $(h^{2,1},h^{1,1})=(11,11)$, and case 2 with fixed points in two twisted sectors as $(h^{2,1},h^{1,1})=(19,19)$. Furthermore, the collapsed cycles $e^g_{ij}$ are actually exceptional cycles which will be considered again in section \ref{sec_3_homology}.

\bigskip
Note that for the $\Z_2\times\Z_2$ orbifold with trivial shifts $\delta_i=0$ the action of for instance $\Theta'$ on the $\Theta$ sector is just multiplication with the discrete torsion $\eta=\pm1$. The resulting Hodge numbers are then $(h^{2,1},h^{1,1})=(3,51)$ and $(h^{2,1},h^{1,1})=(51,3)$ for case of discrete torsion $\eta=+1$ and $\eta=-1$, respectively.


\subsection{Orientifold Background}
\label{sec_2_orientifold_background}

In this subsection, the shifts $\kappa_i$ of the orbifold group \eqref{orb_group} will become important because their values determine the position of the fixed points and the orientifold projection $\Omega\R$ will move them. Therefore, the $\kappa_i$ are no longer set to zero.

\bigskip 
For type IIA string theory, the map $\R$ of the orientifold
\eqref{internal_space} is an anti-holomorphic involution acting on the
internal coordinates $z_i$. In the context of geometric shift orientifolds, the most general form of $\R$ we will use is the following
\begin{equation}\label{orient_group}
  \R : \left\{ \begin{array}{c}
    z_1 \to \bar{z}_1 +\lambda_1\\
    z_2 \to \bar{z}_2 +\lambda_2\\
    z_3 \to \bar{z}_3 +\lambda_3
  \end{array} \right.
\end{equation}
where $\bar{z}$ denotes the complex conjugate of $z$ and the shifts $\lambda_i$ are defined as $\lambda_i=\lambda_i^x\,e^x_i+\lambda_i^y\,e^y_i$ with $\lambda_i^a\in[0,1)$ and $a=x,y$. However, since the point group of $\R$ has to act crystallographically, i.e. lattice points are mapped to lattice points under complex conjugation, one has to require
\begin{equation}
  \beta_i = 0, 1/2 \;.
\end{equation}
Moreover, for $\R$ to be an involution on the torus one has to ensure that
\begin{equation}\begin{array}{l@{\hspace{3pt}}l@{\hspace{40pt}}l}
  \lambda_i^x & {\displaystyle \in\frac{1}{2}\,} \Z
    &\mbox{for~}\beta_i=0 \;, \\
  \lambda_i^x & \in\hphantom{{\displaystyle \frac{1}{2}}\,}\Z
    &{\displaystyle \mbox{for~}\beta_i=\frac{1}{2}} \;.
\end{array}\end{equation}
The closure of the orientifold algebra, i.e. $\left(\Omega\R\cdot\Z_2\times\Z_2\right)\cdot\left(\Omega\R\cdot\Z_2\times\Z_2\right)=\Z_2\times\Z_2$, implies the following restrictions
\begin{align}
  \label{orient_consis_2}
  \beta_i \delta_i^x + \delta_i^y \hspace{28.5pt}
    & \in \frac{1}{2}\,\Z  \,,\\
  \label{orient_consis_3}
  \beta_i \kappa_i^x + \kappa_i^y - \lambda_i^y
    & \in \frac{1}{2}\,\Z  \,,
\end{align}
from which it follows that also fixed points are mapped to fixed points. The explicit mapping of the fixed points under $\R$ is summarized in table \ref{tab_orp_r}.

\begin{table}[htb]
\begin{center}
\begin{tabular}{c|c|c||c|c|c|c}
  $\beta_i$ & $\lambda_i^x$ & $\beta_i\left(\delta_i^x+\kappa_i^x\right)
                               +\left(\delta_i^y+\kappa_i^y\right)
                               -\lambda_i^y\;$ mod $1$ &
  $1\to$ & $2\to$ & $3\to$ & $4\to$ \\
  \hline
  \hline
  $0$           & $0$           & $0$           & $1$ & $2$ & $3$ & $4$ \\
  $0$           & $0$           & $\frac{1}{2}$ & $2$ & $1$ & $4$ & $3$ \\
  $0$           & $\frac{1}{2}$ & $0$           & $3$ & $4$ & $1$ & $2$ \\
  $0$           & $\frac{1}{2}$ & $\frac{1}{2}$ & $4$ & $3$ & $2$ & $1$ \\
  $\frac{1}{2}$ & $0$           & $0$           & $1$ & $2$ & $4$ & $3$ \\
  $\frac{1}{2}$ & $0$           & $\frac{1}{2}$ & $2$ & $1$ & $3$ & $4$
\end{tabular}
\end{center}
\caption{\label{tab_orp_r} Permutation of $\mathbb{I}\,z_i=-z_i+\delta_i+\kappa_i$ fixed points under $\R$. The labels $1\ldots4$ are illustrated in figure \ref{fig_torus}.}
\end{table}

\bigskip
The fixed loci of the orientifold projection $\Omega\R$ are important non-dynamical objects called orientifold planes. On $\T^6=\T^2\times\T^2\times\T^2$, they can be described in terms of the homological basis cycles $[a^i]$ and $[b^i]$ on $\T^2_i$. These cycles are illustrated in figure \ref{fig_torus} and will be introduced more systematically in section \ref{sec_3_homology}. Furthermore, the signs $\eta_{\Omega\R\,g}=\pm1$ indicate the charge of the orientifold planes. We would like to refer the reader not familiar with these concepts to sections \ref{sec_3_homology} and \ref{sec_3_orient_projection}. At this point, the important feature of the expression for the orientifold planes \eqref{orient_plane} are the $\Delta$ symbols.

Working out the fixed set of $\R\,g$ with $g=1,\Theta,\Theta',\Theta\Theta'$ one obtains the orientifold planes in terms of bulk $3$-cycles as follows
\begin{equation}
\begin{array}
{@{}c@{\hspace{-5pt}}r@{\,\cdot\,}c@{\otimes}c@{\otimes}c@{\,\cdot\,}l
 @{\,\cdot\,}l@{}l@{}l}
  [\Pi_{O6}]=& 2\,\eta_{\Omega\R} &
    [a^1]&[a^2]&[a^3] & 1 &
    \Delta_1{\scriptstyle (\lambda_1,0)} &
    \Delta_1{\scriptstyle (\lambda_2,0)} &
    \Delta_1{\scriptstyle (\lambda_3,0)} \\
  & -2\,\eta_{\Omega\R\Theta} & [b^1]&[b^2]&[a^3] & 2^{-2\beta_1-2\beta_2} &
    \Delta_2{\scriptstyle (\lambda_1,\kappa_1) } &
    \Delta_2{\scriptstyle (\lambda_2,\kappa_2) } &
    \Delta_1{\scriptstyle (\lambda_3,\delta_3) } \\
  & -2\,\eta_{\Omega\R\Theta'} & [a^1]&[b^2]&[b^3] & 2^{-2\beta_2-2\beta_3} &
    \Delta_1{\scriptstyle (\lambda_1,\delta_1) } &
    \Delta_2{\scriptstyle (\lambda_2,\delta_2+\kappa_2) } &
    \Delta_2{\scriptstyle (\lambda_3,\kappa_3) } \\
  & -2\,\eta_{\Omega\R\Theta\Theta'} & [b^1]&[a^2]&[b^3]
    & 2^{-2\beta_1-2\beta_3} &
    \Delta_2{\scriptstyle (\lambda_1,\delta_1+\kappa_1) } &
    \Delta_1{\scriptstyle (\lambda_2,\delta_2) } &
    \Delta_2{\scriptstyle (\lambda_3,\delta_3+\kappa_3) }
\end{array}\label{orient_plane}
\end{equation}
where the symbols $\Delta_1$ and $\Delta_2$ are defined as
\begin{equation}\label{orient_plane_delta}
\begin{array}{lcl}
  \Delta_1(\lambda_i,\delta_i) & = & \left\{
    \begin{array}{ll}
    1 & \hspace{5pt}\mbox{if}\hspace{132pt}\lambda_i^x+\delta_i^x \in\Z\\
    0 & \hspace{5pt}\mbox{otherwise}
    \end{array}
    \right. \,,\\[30pt]
  \Delta_2(\lambda_i,\delta_i+\kappa_i) & = & \left\{
    \begin{array}{ll}
    1 & \begin{array}{llcl}
        \mbox{if}    & \beta_i=0           & \mbox{and} &
        -\lambda^y_i+\delta_i^y+\kappa_i^y\in\Z \\
        \mbox{or if} & \beta_i=\frac{1}{2}
        \end{array} \\[15pt]
    0 & \hspace{5pt}\mbox{otherwise}
    \end{array}
    \right.\;.
\end{array}\end{equation}
Note that, depending on the complex structures $\beta_i$ and the various shifts $\delta_i$, $\kappa_i$, $\lambda_i$, the orientifold plane can receive contributions from zero, one two, three or four sectors. This is in contrast to the usual $\Z_2\times\Z_2$ orientifold with always four contributions.


\subsection{Summary}

Let us summarize the considerations so far. The general action of the shift $\Z_2\times\Z_2$ orbifold group is displayed in equation \eqref{orb_group} and the action of the orientifold projection is shown in \eqref{orient_group}. In these expressions there appear three types of shifts called of type I, II and III in the following. Their definitions and consistency conditions are
\begin{center}
\begin{tabular}{@{}l@{\;:\quad}l@{\,,\quad}l@{\,,\quad}
r@{\hspace{2pt}}c@{\hspace{2pt}}l@{}}
  type I shifts & $\delta_i = \delta_i^x\,e_i^x+\delta_i^y\,e^y_i$ &
    $\delta_i^a=0,\frac{1}{2}$ & $\beta_i\delta_i^x+\delta_i^y\hspace{28.5pt}$
    &$\in$&$
    \frac{1}{2}\Z$, \\
  type II shifts & $\kappa_i = \kappa_i^x\,e_i^x+\kappa_i^y\,e^y_i$ &
    $\kappa_i^a\in[0,1)$ & $\beta_i\kappa_i^x+\kappa_i^y-\lambda_i^y
    $&$\in$&$\frac{1}{2}\Z$, \\
  type III shifts & $\lambda_i = \lambda_i^x\,e_i^x+\lambda_i^y\,e^y_i$ &
    $\lambda_i^a\in[0,1)$ & $\lambda_i^x$&$\in$&$
    \left(\frac{1}{2}+\beta_i\right)\Z$.
\end{tabular}
\end{center}

The type I shifts determine the Hodge numbers of the orbifold and therefore the number of homological cycles. For three, two and one non-trivial shifts $\delta_i$ one respectively finds configurations with $(h^{2,1},h^{1,1})=(3,3)$, $(11,11)$ and $(19,19)$. In the case of all $\delta_i=0$, one obtains Hodge numbers $(h^{2,1},h^{1,1})=(3,51)$ and $(51,3)$ depending on the discrete torsion $\eta$.

The type II and type III shifts become important at the level of the orientifold. Their values determine the mapping of the fixed points as one can see from table \ref{tab_orp_r}. Moreover, $\kappa_i$ and $\lambda_i$ also specify the fixed set under the orientifold projection $\Omega\R$, and therefore also the orientifold planes. This in turn implies, that the type II and type III shifts strongly influence the Ramond-Ramond tadpole cancellation condition.

\bigskip
In order to illustrate this point, let us consider topologies with Hodge numbers $(3,51)$ or $(51,3)$ and the following combination of complex structures and shifts
\begin{equation}\begin{array}
{l@{\hspace{30pt}}l@{\hspace{30pt}}ll@{\hspace{30pt}}l}
  \beta_1=0, & \delta_1=0, & \kappa_1=0,             && \lambda_1=0, \\
  \beta_2=0, & \delta_2=0, & \kappa_2=0,             && \lambda_2=0, \\
  \beta_3=0, & \delta_3=0, & \kappa^x_3=0,
                           & \kappa_3^y=\frac{1}{2},  & \lambda_3=0.
\end{array}\end{equation}
In contrast to the four sectors of the orientifold plane for the standard $\Z_2\times\Z_2$ orientifold, the resulting expression in this case has only two contributions
\begin{equation}\begin{array}
{c@{\hspace{3pt}}r@{\hspace{3pt}}l@{\,\cdot\,}c@{\otimes}c@{\otimes}cl}
  [\Pi_{O6}]=& 2&\eta_{\Omega\R} &  [a^1]&[a^2]&[a^3] \\
  & -2&\eta_{\Omega\R\Theta} & [b^1]&[b^2]&[a^3] &.
\end{array}\end{equation}
As a second example let us consider a configuration with Hodge numbers $(3,3)$ and shifts and complex structures chosen as
\begin{equation}\begin{array}
{l@{\hspace{30pt}}ll@{\hspace{30pt}}l@{\hspace{30pt}}l}
  \beta_1=\frac{1}{2}, & \delta^x_1=0, & \delta^y_1=\frac{1}{2}, & \kappa_1=0,
    & \lambda_1=0, \\
  \beta_2=\frac{1}{2}, & \delta^x_2=0, & \delta^y_2=\frac{1}{2}, & \kappa_2=0,
    & \lambda_2=0, \\
  \beta_3=\frac{1}{2}, & \delta^x_3=0, & \delta^y_3=\frac{1}{2}, & \kappa_3=0,
    & \lambda_3=0.
\end{array}\end{equation}
Usually one would expect that there is only one sector contributing to the orientifold plane. However, in this case there are all four sectors present which reads as
\begin{equation}\begin{array}
{c@{\hspace{3pt}}r@{\hspace{3pt}}l@{\,\cdot\,}c@{\otimes}c@{\otimes}cl}
  [\Pi_{O6}]=& 2&\eta_{\Omega\R} &  [a^1]&[a^2]&[a^3] \\
  & -\frac{1}{2}&\eta_{\Omega\R\Theta} & [b^1]&[b^2]&[a^3] \\
  & -\frac{1}{2}&\eta_{\Omega\R\Theta'} & [a^1]&[b^2]&[b^3]  \\
  & -\frac{1}{2}&\eta_{\Omega\R\Theta\Theta'} & [b^1]&[a^2]&[b^3] &.
\end{array}\end{equation}


\section{Rigid Branes on Shift $Z_2\times Z_2$ Orientifolds}
\label{sec_3}

In this section, the necessary techniques to analyze intersecting D-branes models on shift $\Z_2\times\Z_2$ orientifolds are developed. The branes under consideration are supersymmetric D6-branes filling out $4$-dimensional space-time and three dimensions of the internal space.

\bigskip
In order to fix the open-string moduli, we are going to construct rigid cycles and fractional branes as in \cite{Blumenhagen:2005tn}. Therefore, not all configurations of shift $\Z_2\times\Z_2$ orientifolds will be of interest to us.
\begin{itemize}
\itemsep=0pt\parskip=0pt
\item The standard $\Z_2\times\Z_2$ orientifold without discrete torsion, i.e. with Hodge numbers $(h^{2,1},h^{1,1})=(3,51)$, does not admit rigid $3$-cycles. Therefore it will not be considered in this work.

\item The standard $\Z_2\times\Z_2$ orientifold with discrete torsion, i.e. with Hodge numbers $(h^{2,1},h^{1,1})=(51,3)$ was analyzed in \cite{Blumenhagen:2005tn} for trivial type II and type III shifts $\kappa_i=\lambda_i=0$. These constructions can easily be generalized to the cases with $\kappa_i\neq0$ and $\lambda_i\neq0$ and will therefore not be considered further in this work.

\item The shift $\Z_2\times\Z_2$ orientifold with all $\delta_i\neq0$, i.e. with Hodge numbers $(h^{2,1},h^{1,1})=(3,3)$, does not admit rigid cycles and therefore will not be considered.

\item The shift $\Z_2\times\Z_2$ orientifolds with Hodge numbers $(h^{2,1},h^{1,1})=(11,11)$ and $(19,19)$ allow for rigid cycles and will be analyzed in detail in the following.
\end{itemize}


\subsection{Homology}
\label{sec_3_homology}


\subsubsection*{Bulk Cycles}

For the standard $\T^2_i$ there are two $1$-cycles which wrap exactly once around the torus. Their homology classes are denoted by $[a^i]$ and $[b^i]$ and the homological intersection numbers are defined as $[a^i]\circ[b^j]=-\delta^{ij}$ and all others vanishing. The two different complex structures $\beta_i=0$ and $\beta_i=1/2$ can be combined in defining $[{a'}^i]=[a^i]+\beta_i\,[b^i]$ which is illustrated in figure \ref{fig_torus}.

The factorizable $3$-cycles on $\T^2\times\T^2\times\T^2$ are obtained as the direct product of $1$-cycles from each $\T^2$ as follows
\begin{equation}
  [\Pi^{T^6}_a]=\bigotimes_{i=1}^3\left(n^i_a\,[{a'}^i]
    + m^i_a\,[b^i]\right) = \bigotimes_{i=1}^3\left(n^i_a\,[a^i]
    + \tilde{m}^i_a\,[b^i]\right) \,.
\end{equation}
Here $n^i_a$ and $\tilde{m}_a^i=m^i_a +\beta_i\,n_a^i$ denote the wrapping
numbers of the $1$-cycles in each $\T^2$ factor and the label $\T^6$ indicates
that these cycles are from the background torus.

\bigskip
As a next step let us consider the orbifold. For the factorizable $3$-cycles this means that one includes all the images of $[\Pi_a^{T^6}]$ under the orbifold group
\begin{equation}
  [\Pi_a^B] = \sum_{g\in \Z_2\times \Z_2}[ g\:\Pi_a^{T^6}]
    =4\:[\Pi_a^{T^6}] \,.
\end{equation}
Then the orbifold identification will be such that one has to take into account
appropriate factors in the calculation of intersection numbers. The orbifold cycles coming from the ambient space are usually called bulk-cycles which is indicated by the superscript $B$. The homological intersection numbers between different bulk-cycles is computed as
\begin{equation}
  [\Pi_a^B]\circ[\Pi_b^B]
    = \frac{1}{4}\;4\:[\Pi_a^{T^6}]\circ 4\:[\Pi_b^{T^6}]
    = 4\;\prod_{i=1}^3\left(n_a^i\tilde{m}_b^i-m_a^i\tilde{n}_b^i\right)\,,
\end{equation}
where the factor $1/4$ accounts for the identification of cycles and intersection points under the orbifold group. For later convenience the intersection number on the i'th torus will be denoted as $I_{ab}^i=n_a^i\tilde{m}_b^i-\tilde{m}_a^i n_b^i$. The intersection number on a factorizable $\T^6$ then is $I_{ab}=I_{ab}^1\cdot I_{ab}^2\cdot I_{ab}^3$.


\subsubsection*{Exceptional Cycles}

Let us now turn to the exceptional cycles coming from the collapsed $\mathbb{P}^1$s. These $2$-cycles will be denoted as $e_{ij}^g$ where the $i,j$ are in the range from $1$ to $4$ and label the fixed points as in figure \ref{fig_torus}. The label $g=\Theta,\Theta',\Theta\Theta'$ indicates the twisted sector and the intersection number can be found as
\begin{equation}\label{cycles_in1}
  [ e_{ij}^g]\circ[ e_{uv}^h] = -2\,\delta_{iu}\,\delta_{jv}\,\delta^{gh}
\end{equation}
where the $\delta$ are Kronecker $\delta$. However, the orbifold group will act on the exceptional $2$-cycles. The action of non-trivial $h\neq g$ can be with ($\eta=-1$) or without ($\eta=+1$) discrete torsion. Furthermore, $h$ can also permute the fixed points as
\begin{equation}\label{orb_pq}
  g\:e_{ij}^h = \eta\,e_{p(i)q(j)}^h
\end{equation}
where $g,h\in \Z_2\times\Z_2$, $g\neq h$ and $g,h\neq 1$. The order two permutations $p$ and $q$ depend only on the shift $\delta_i$ (not on $\kappa_i$ or $\lambda_i$) in the corresponding torus and their action is listed in table \ref{tab_cycles1}.
\begin{table}[ht]
\begin{center}
\begin{tabular}{c||c|c|c|c}
  $p,q$   & $\delta_i=0$ & $\delta_i=1/2\,e_i^x$ & $\delta_i=1/2\,e_i^y$
          & $\delta_i=1/2\,(e_i^x+e_i^y)$ \\ \hline \hline
  1 $\to$ & 1 & 3 & 2 & 4 \\
  2 $\to$ & 2 & 4 & 1 & 3 \\
  3 $\to$ & 3 & 1 & 4 & 2 \\
  4 $\to$ & 4 & 2 & 3 & 1
\end{tabular}
\end{center}
\caption{\label{tab_cycles1} Permutations of the fixed points in $\T^2_i$ under the orbifold group depending on the type I shift $\delta_i$.}
\end{table}


\subsubsection*{The Case with Hodge Numbers (11,11)}

In the following, the configuration with fixed points only in the $\Theta$ twisted sector will be chosen for the case with Hodge numbers $(11,11)$. The exceptional $2$-cycles $e_{ij}^{\Theta}$ can be combined with a bulk $1$-cycle from $\T^2_3$ to obtain an exceptional $3$-cycle. As an $\Z_2\times\Z_2$ invariant basis one finds
\begin{equation}\begin{split}
  [\alpha^{\Theta}_{ij,n}]\mspace{3.7mu} &= \left([e^{\Theta}_{ij}]
    -\eta\,[e^{\Theta}_{p(i)q(j)}]\right)\otimes[a^3]\,, \\
  [\alpha^{\Theta}_{ij,m}] &= \left([e^{\Theta}_{ij}]
    -\eta\,[e^{\Theta}_{p(i)q(j)}]\right)\otimes[b^3]\,,
\end{split}\end{equation}
where $i,j$ denote the fixed points in the first and second torus, respectively. With this basis one can then build more general exceptional $3$-cycles as
\begin{equation}
  [\Pi^{\Theta}_{a,ij}]=n_a^3\,[\alpha^{\Theta}_{ij,n}]
  +\tilde{m}_a^3\,[\alpha^{\Theta}_{ij,m}].
\end{equation}
The intersection number between those cycles can be calculated using \eqref{cycles_in1} and one obtains
\begin{equation}
  [\Pi^{\Theta}_{a,ij}]\circ[\Pi^{\Theta}_{b,st}]
  = 2\, I_{ab}^3 \,\left( \delta_{is}\delta_{jt}-\eta
  \,\delta_{p(i)s}\delta_{q(j)t}\right)
\end{equation}
where the identification under the remaining $\Z_2$ has also been taken into account.


\subsubsection*{The Case with Hodge Numbers (19,19)}

For the case with Hodge numbers $(19,19)$ the configuration with fixed points in the $\Theta$ and $\Theta'$ sector will be chosen in the following. As an invariant basis for the exceptional $3$-cycles one finds
\begin{equation}\begin{array}{lcl}
  [\alpha^{\Theta}_{ij,n}]&=&\bigl([e^{\Theta}_{ij}]-
    \eta\,[e^{\Theta}_{iq(j)}]\bigr)\otimes[a^3]\,, \\ \vphantom{x}
  [\alpha^{\Theta}_{ij,m}]&=&\bigl([e^{\Theta}_{ij}]-
    \eta\,[e^{\Theta}_{iq(j)}]\bigr)\otimes[b^3]\,, \\ \vphantom{x}
  [\alpha^{\Theta'}_{kl,n}]&=&
    \bigl([e^{\Theta'}_{kl}]-\eta\,[e^{\Theta'}_{q(k)l}
    ]\bigr)\otimes [a^1]\,, \\ \vphantom{x}
  [\alpha^{\Theta'}_{kl,m}] &=&
    \bigl([e^{\Theta'}_{kl}]-\eta\,[e^{\Theta'}_{q(k)l}
    ]\bigr)\otimes[b^1] \,,
\end{array}\end{equation}
where $p$ and $q$ are the permutations introduced in equation \eqref{orb_pq}, $i,j$ denote the fixed points in the first and second $\T^2$ for the
$\Theta$ sector, and $k,l$ denote the fixed points for the second and third
$\T^2$ in the $\Theta'$ sector. General exceptional $3$-cycles can be constructed as
\begin{equation}\begin{array}{lcl}
  [ \Pi^{\Theta}_{a,ij} ] & = & n_a^3\, [ \alpha^{\Theta}_{ij,n} ]
  +\tilde{m}_a^3\, [ \alpha^{\Theta}_{ij,m} ]\,, \\
  \vphantom{x}[ \Pi^{\Theta'}_{a,kl} ] &=& n_a^1\,[\alpha^{\Theta'}_{kl,n}]
  +\tilde{m}_a^1\, [ \alpha^{\Theta'}_{kl,m} ]\,,
\end{array}\end{equation}
and the only non-vanishing intersection numbers of these cycles are
\begin{equation}\begin{array}{ccl}
  [\Pi^{\Theta}_{a,ij}]\circ[\Pi^{\Theta}_{b,st}]
  &=&2\,I_{ab}^3\,\delta_{is}\,\left(\delta_{jt}
    -\eta\,\delta_{q(j)t}\right)\,, \\
  \vphantom{x}[\Pi^{\Theta'}_{a,kl}]\circ[\Pi^{\Theta'}_{b,uv}]
  &=&2\,I_{ab}^1\,\delta_{lv}\,\left(\delta_{ku}
    -\eta\,\delta_{q(k)u}\right) \; .
\end{array}\end{equation}


\subsection{Fractional Branes}
\label{sec_3_frac_branes}

After constructing the bulk and exceptional cycles on the orbifold, one can now
combine them to build fractional cycles. In order to fix the open-string moduli of a brane, it will be interesting if complete rigid cycles appear.


\subsubsection*{Fixed Points}

From a geometrical point of view it is clear that the exceptional cycles $e^g_{a,ij}$ can be turned on only if the bulk-brane runs through the corresponding fixed points. Therefore, in order to obtain rigid cycles, we place the brane such that it runs through at least one of them. However, since the branes of interest are supersymmetric and hence special Lagrangian,\footnote{See part 2 of section \ref{sec_3_consistency} for a detailed explanation.} they are straight lines in each $\T^2$ and this implies that they run through exactly two fixed points. Depending on the bulk wrapping numbers $(n_a^i,m_a^i)$, only certain combinations of fixed points are allowed. These are summarized in table \ref{tab_frac_fp}.\footnote{Note that for the case of Hodge numbers $(19,19)$ there are further restrictions on the allowed fixed points in $\T^2_2$ depending on the wrapping numbers $(n_a^2,m_a^2)$. However, these relations are needed only on a technical level.}
\begin{table}[ht]
\begin{center}
\begin{tabular}{c||c}
  $(n^i,m^i)$& fixed points in $\T^2_i$ \\ \hline \hline
  (odd,odd)  & \{1,4\} or \{2,3\}  \\
  (odd,even) & \{1,3\} or \{2,4\}  \\
  (even,odd) & \{1,2\} or \{3,4\}
\end{tabular}
\end{center}
\caption{Wrapping numbers and allowed fixed points for one $\T_i^2$.}
\label{tab_frac_fp}
\end{table}

\bigskip
Let us now define a fixed point set $S^g_a$ for brane $a$ in the $g$ twisted sector. The fixed points coming from $\Theta$ will be labelled as $\{i_1,i_2\}$ in $\T^2_1$ and as $\{j_1,j_2\}$ in $\T^2_2$. For the $\Theta'$ sector the labels $k$ in $\T^2_2$ and $l$ in $\T^2_3$ will be used
\begin{equation}\begin{split}
  S_a^{\Theta} =& \left\{\{i_1,i_2\},\{j_1,j_2\}\right\}\,, \\
  S_a^{\Theta'}=& \left\{\{k_1,k_2\},\{l_1,l_2\}\right\}\,.
\end{split}\end{equation}
As mentioned previously, depending on the shift $\delta_i$ in a two-torus the orbifold group will permute the fixed points of one sector. In particular, also the fixed point sets $S^g_a$ are changed. However, in certain cases this action just interchanges the fixed points in $S^g_a$ as $p(i_1)=i_2$ and $p(i_2)=i_1$ which means that the bulk-brane through such points is fixed under the group action. The conditions for this interchange are summarized in table \ref{tab_frac_star} and wrapping numbers leading to such a fixed point set will be labelled by a star ($\star$) in the following. Note that for non-star wrapping numbers $\{i_1,i_2\}$ and $\{p(i_1),p(i_2)\}$ are complementary.
\begin{table}[ht]
\begin{center}
\begin{tabular}{c||c}
  $\delta_i$           & $(n^i,m^i)$ \\ \hline \hline
  $1/2\;(e^x_i+e^y_i)$ & (odd,odd)   \\
  $1/2\;e^x_i$         & (odd,even)  \\
  $1/2\;e^y_i$         & (even,odd)  \\
\end{tabular}
\end{center}
\caption{Bulk-cycle wrapping numbers on $\T^2_i$ which lead to an invariant fixed point set (up to permutations $p(i_1)=i_2$) for a given group action $\delta_i$.} \label{tab_frac_star}
\end{table}


\subsubsection*{Charges}

From a geometrical point of view, there are two possibilities for an exceptional $2$-cycle to run through the blown-up fixed point $\mathbb{P}^1$.
These two orientations will be denoted as $\epsilon_{a,ij}^g=\pm1$ where $i,j$
label again the fixed points and $g$ indicates the twisted sector. However, the
$\epsilon_{a,ij}^g$ can be changed as follows,
\begin{align}\begin{split}\begin{array}{rcl}
  (n_a^1,\tilde{m}_a^1)\otimes(n_a^2,\tilde{m}_a^2)\otimes
    (n_a^3,\tilde{m}_a^3)&\to&(+ n_a^1,+\tilde{m}_a^1)\otimes(-
    n_a^2,-\tilde{m}_a^2)\otimes(-n_a^3,-\tilde{m}_a^3)  \\
  \Rightarrow\qquad\epsilon_{a,ij}^{\Theta} &\to&
    -\epsilon_{a,ij}^{\Theta} \\
  \Rightarrow\qquad\epsilon_{a,ij}^{\Theta'} &\to&
    +\epsilon_{a,ij}^{\Theta'} \,,\\[3mm]
  (n_a^1,\tilde{m}_a^1)\otimes(n_a^2,\tilde{m}_a^2)\otimes
    (n_a^3,\tilde{m}_a^3)&\to&(-n_a^1,-\tilde{m}_a^1)\otimes(-
    n_a^2,-\tilde{m}_a^2)\otimes(+ n_a^3,+\tilde{m}_a^3) \\
  \Rightarrow\qquad\epsilon_{a,ij}^{\Theta} &\to&
    +\epsilon_{a,ij}^{\Theta} \\
  \Rightarrow\qquad\epsilon_{a,ij}^{\Theta'} &\to&
    -\epsilon_{a,ij}^{\Theta'} \, .
\end{array}\end{split}\end{align}
Since the bulk-brane is invariant under such a change of wrapping numbers, one
can use this redundancy to fix signs as $\epsilon_{a,i_1j_1}^{\Theta}=+1$ and $\epsilon_{a,k_1l_1}^{\Theta'}=+1$.

\begin{figure}[p]\begin{center}
\includegraphics[totalheight=0.19\textheight]{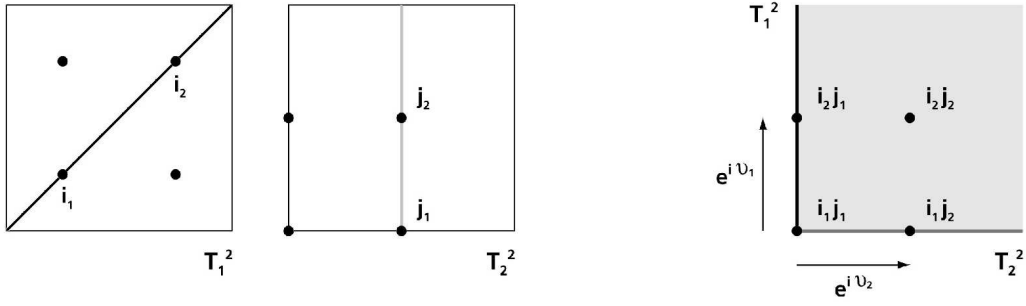}
\caption{On the left one can see a brane which runs through two fixed points in each $\T^2$ factor and on the right the direct product of the brane in $\T^2_1\times\T^2_2$ is shown. The $\vartheta_{1,2}$ correspond to Wilson lines along the brane.\label{fig_charges}}
\end{center}\end{figure}

\bigskip
This geometric picture also has a physical meaning. The orientations
$\epsilon^g_{a,ij}$ can be regarded as the charges of the Ramond-Ramond (R-R) fields localized at the $\Z_2\times\Z_2$ fixed points. These charges correspond to turning on discrete Wilson lines along the brane as it was shown in \cite{Sen:1998ii} and as illustrated in figure \ref{fig_charges}. The resulting relations for
the charges are then
\begin{center}\begin{tabular}
{>{$}l<{$}>{$}c<{$}>{$}l<{$}
 >{$}c<{$}>{$}l<{$}>{$}c<{$}>{$}l<{$}}
  \epsilon^{\Theta}_{a,i_1j_1} &=&
    +1, &\hphantom{\hspace{30pt}}& \epsilon^{\Theta'}_{a,k_1l_1} &=&
    +1, \\
  \epsilon^{\Theta}_{a,i_2j_1} &=&
    +1\cdot e^{i\vartheta_1}, && \epsilon^{\Theta'}_{a,k_2l_1} &=&
    +1\cdot e^{i\vartheta_2}, \\
  \epsilon^{\Theta}_{a,i_1j_2} &=&
    +1\cdot e^{i\vartheta_2}, && \epsilon^{\Theta'}_{a,k_1l_2} &=&
    +1\cdot e^{i\vartheta_3}, \\
  \epsilon^{\Theta}_{a,i_2j_2} &=&
    +1\cdot e^{i\vartheta_1}e^{i\vartheta_2}, &&
    \epsilon^{\Theta'}_{a,k_2l_2} &=& +1\cdot e^{i\vartheta_2}e^{i\vartheta_3},
\end{tabular}\end{center}
where the above redundancy was used to fix $\epsilon_{a,i_1j_1}^{\Theta}$ and $\epsilon_{a,k_1l_1}^{\Theta'}$ as $+1$ and the $\vartheta_{1,2,3}=0,\pi$ are the possible Wilson lines along the brane in $\T^2_1$, $\T^2_2$ and $\T^2_3$, respectively. With these relations it is easy to see that the charges satisfy the following equation \cite{Sen:1998ii}
\begin{equation}
  \sum_{i,j\in S_a^g} \epsilon^g_{a,ij}=0\mbox{~mod~} 4
\end{equation}
for each $g$. If this condition is not satisfied it would correspond to turning on a constant magnetic flux on the D6-brane which is not considered in this work. The general result for the charges $\epsilon_{a,ij}^g$ is shown in table
\ref{tab_frac_wil}.
\begin{table}[ht]
\begin{center}
\begin{tabular}{c||cc}
        & $j_1$   & $j_2$          \\ \hline\hline
  $i_1$ & $+1$    & $\iota'$       \\
  $i_2$ & $\iota$ & $\iota\iota'$
\end{tabular}
\hspace{2cm}
\begin{tabular}{c||cc}
        & $l_1$   & $l_2$          \\ \hline\hline
  $k_1$ & $+1$    & $\iota''$      \\
  $k_2$ & $\iota'$& $\iota'\iota''$
\end{tabular}
\end{center}
\caption{Inequivalent choices for the signs $\epsilon^{\Theta}_{a,ij}$ and
$\epsilon^{\Theta'}_{a,kl}$ where $\iota,\iota',\iota''=\pm1$.}
\label{tab_frac_wil}
\end{table}


\begin{figure}[p]\begin{center}
\includegraphics[totalheight=0.175\textheight]{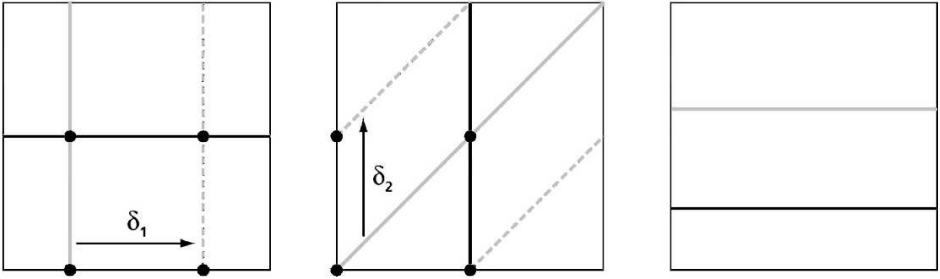}
\caption{Fractional branes for the case of Hodge numbers $(11,11)$. The type I shifts are $\delta=\frac{1}{2}(e_1^x+e_2^y)$ as indicated. The solid grey line is a fractional brane of type $\Pi_a^{F11}$ and its bulk-brane orbifold image is indicated as a dotted line. The black line is a fractional brane of type star ($\Pi_{\star\,b}^{F11}$) which is its own bulk-brane orbifold image.\label{fig_frac11}}
\end{center}\end{figure}

\begin{figure}[p]\begin{center}
\includegraphics[totalheight=0.175\textheight]{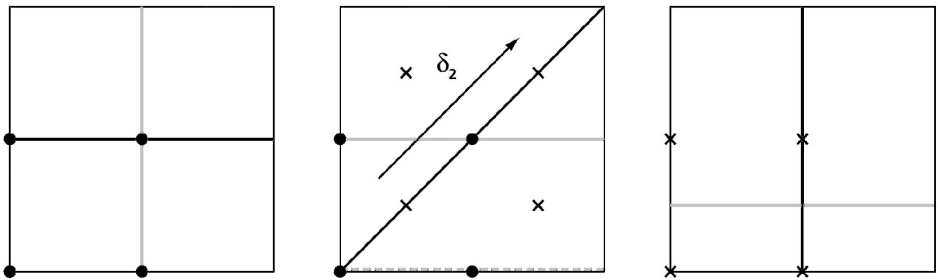}
\caption{Example for the case of Hodge numbers $(19,19)$. The type I shift is
$\delta=e_2^x+e_2^y$ as indicated. A fractional brane of type
$\Pi_{\Theta}^{F19}$ is shown as a grey line and a fractional brane of type star ($\Pi_{\star\,b}^{F19}$) is drawn as a black line.\label{fig_frac19}}
\end{center}\end{figure}



\subsubsection*{The Case with Hodge Numbers (11,11)}

Let us now turn to the explicit construction of fractional branes. For the 
case with Hodge numbers $(11,11)$ we make the following general ansatz
\begin{equation}
  \Pi^{F11}_a = \frac{1}{N^B_a}\Pi^B_a + \frac{1}{N^{(1)}_a}\sum_{i,j\in
  S^{\Theta}_a}\epsilon^{\Theta}_{a,ij} \Pi^{\Theta}_{a,ij}
\end{equation}
where the normalization constants $N^B_a$ and $N^{(1)}_a$ can be 
determined from the number of fractional branes needed to obtain a bulk-brane
\begin{equation}
  N^B_a = N^{(1)}_a = 2 \;.
\end{equation}
However, there is a subtlety for the branes of type star ($\star$), i.e. for those with $i_2=p(i_1)$ and $j_2=q(j_1)$. The exceptional part of these branes is invariant under the orbifold group only if the fixed point charges satisfy
\begin{equation}\label{frac_consistency_1}
  \iota_a \iota_a' = -\eta
\end{equation}
where $\eta$ is the discrete torsion. Furthermore, for the group action on 
the open-string lattice states to be well-defined, equation
\eqref{frac_consistency_1} has to be satisfied for all branes $a$ in a 
specific D-brane configuration.

In summary, the fractional branes in the case with Hodge numbers $(11,11)$ are the following
\begin{equation}
\label{fraca}
\begin{split}
  \Pi_a^{F11}&= \frac{1}{2}\Pi^B_a + \frac{1}{2}\sum\nolimits_{\{i,j\}\in
    S^{\Theta}_a}\epsilon^{\Theta}_{a,ij} \Pi^{\Theta}_{a,ij}\,, \\
  \Pi_{\star\,a}^{F11}&= \frac{1}{2}\Pi^B_a + \hphantom{\frac{1}{2}}\left(
    \Pi^{\Theta}_{a,i_1j_1}+\iota_a\Pi^{\Theta}_{a,i_2j_1}\right) \,,
\end{split}\end{equation}
where for the star ($\star$) branes the exceptional part is worked out explicitly. Two examples for these fractional branes in the case of Hodge numbers $(11,11)$ are presented in figure \ref{fig_frac11}.

As expected, from \eqref{fraca} one can see that there are no complete rigid branes. The reason is that there is one twisted sector with fixed points only in the first and second $\T^2$. The bulk-brane in $\T^2_3$ is not fixed and therefore a chiral multiplet transforming in the adjoint representation will appear in the spectrum.


\subsubsection*{The Case with Hodge Numbers (19,19)}

For the $(19,19)$ configuration we make a similar ansatz as above with two 
twisted sectors
\begin{equation}
  \Pi_a^{F19} = \frac{1}{N^B_a}\Pi^B_a + \frac{1}{N^{(1)}_a}\sum_{i,j\in
  S^{\Theta}_a}\epsilon^{\Theta}_{a,ij} \Pi^{\Theta}_{a,ij} +
  \frac{1}{N^{(2)}_a}\sum_{k,l\in
   S^{\Theta'}_a}\epsilon^{\Theta'}_{a,kl} \Pi^{\Theta'}_{a,kl}\,.
\end{equation}
Remember that only those exceptional cycles are turned where the bulk-brane runs through. Then there are three cases: the $\Theta$ sector is turned on, the $\Theta'$ sector is turned on or both sectors are turned on. If a sector $g$ is not present then the corresponding fixed point set will be treated as $S^g_a=\varnothing$.

Again, there is a subtlety for the branes of type star ($\star$), i.e. for
those with fixed points $j_2=q(j_1)$ in $\T^2_2$. To have an
exceptional 
part invariant under the orbifold group, one has to require that
\begin{equation}\label{frac_consistency_2}
  \iota_a' = -\eta
\end{equation}
where $\eta$ is the discrete torsion. Moreover, for a well-defined action
on the open-string ground states, \eqref{frac_consistency_2} has to be
satisfied for all branes $a$ in a specific D-brane configuration.

The normalization constants $N^B_a$, $N^{(1)}_a$ and $N^{(2)}_a$ can be 
determined from the number of fractional branes needed to obtain a 
bulk-brane. However, in contrast to the $(11,11)$ case, this time there is
a difference between the star ($\star$) and non-star branes
\begin{equation}
  N^B_a          = N^{(1,2)}_a          = 2\;, \hspace{40pt}
  N^B_{\star\,a} = N^{(1,2)}_{\star\,a} = 4\;.
\end{equation}
In summary, the fractional branes for the case with Hodge numbers $(19,19)$ are the following
\begin{equation}\label{frac_19}\begin{split}
  \Pi^{F19}_{\Theta\,a}&= \frac{1}{2}\Pi^B_a + \frac{1}{2}\sum\nolimits_{i,j
    \in S^{\Theta}_a}\epsilon^{\Theta}_{a,ij} \Pi^{\Theta}_{a,ij}\,, \\
  \Pi^{F19}_{\Theta'\,a}&= \frac{1}{2}\Pi^B_a + \frac{1}{2}\sum\nolimits_{k,l
    \in S^{\Theta'}_a}\epsilon^{\Theta'}_{a,kl} \Pi^{\Theta'}_{a,kl}\,, \\[10pt]
  \Pi^{F19}_{\star\,a}&= \frac{1}{4}\Pi^B_a + \frac{1}{2}\bigl(
    \Pi^{\Theta}_{a,i_1j_1} + \iota_a \Pi^{\Theta}_{a,i_2j_1} \bigr)
    +\frac{1}{2}\bigl( \Pi^{\Theta'}_{a,k_1l_1}
    + \iota''_a \Pi^{\Theta'}_{a,k_1l_2} \bigr)\,,
\end{split}\end{equation}
where $\Theta$ labels a fractional brane with fixed points in the $\Theta$
sector, $\Theta'$ labels the $\Theta'$ sector and a star ($\star$) labels a
fractional brane with fixed points in both sectors. Note that the 
exceptional part in the last line is worked out explicitly and that two 
examples for fractional branes can be found in figure \ref{fig_frac19}.

One can see that the first two types of branes in \eqref{frac_19} are not completely rigid and a chiral multiplet transforming in the adjoint representation will arise. But the last type of branes is completely rigid and therefore no open string moduli fields will appear.

\bigskip
However, for the case with Hodge numbers$(19,19)$ there exists an 
interesting linear 
combination of fractional branes. Consider two branes of type star ($\star$) with wrapping numbers $(n^1,m^1)\otimes(n^2,m^2)\otimes(n^3,m^3)$ and $(-n^1,-m^1)\otimes(n^2,m^2)\otimes(-n^3,-m^3)$. In a linear combination the fixed point charges cancel and therefore the exceptional part vanishes
\begin{equation}\label{half_bulk_brane}\begin{split}
  \Pi^{F19}_{\star\,a}\left( \scriptstyle{(n^1,m^1)\otimes(n^2,m^2)
  \otimes(n^3,m^3)} \right) & \\
  + \Pi^{F19}_{\star\,a}\left( \scriptstyle{(-n^1,-m^1)\otimes(n^2,m^2)
  \otimes(-n^3,-m^3)} \right) &= \frac{1}{2} \Pi_a^B \;.
\end{split}\end{equation}
This half bulk-brane is no longer fixed in the second $\T^2$ and an 
open-string moduli field appears, but in the first and third two-torus the 
brane remains rigid.


\subsubsection*{Summary}

At this point, it is worth to summarize the constructions of fractional 
branes so far. For the case of Hodge numbers $(11,11)$ there are no complete rigid branes because there are fixed points only in two $\T^2$ factors. However, for the case of Hodge numbers $(19,19)$ the branes of type star ($\star$) are completely rigid.

\bigskip
A brane of type star ($\star$) in the case of Hodge numbers $(19,19)$ has the property that the fixed point sets $S^{\Theta}=\{\{i_1,i_2\},\{j_1,j_2\}\}$ and $S^{\Theta'}=\{\{k_1,k_2\},\{l_1,l_2\}\}$ are invariant under the action of the orbifold group up to permutations. That means that
\begin{equation}\begin{split}
  q(j_1)=j_2\;, & \hspace{60pt} q(k_1)=k_2 \;,\\
  q(j_2)=j_1\;, & \hspace{60pt} q(k_2)=k_1 \;,
\end{split}\end{equation}
where $q$ denotes the permutation of fixed points displayed in table \ref{tab_frac_fp}. From a geometrical point of view, such a brane runs through fixed points of the $\Theta$ {\em and} $\Theta'$ sector and its bulk-brane is mapped to itself under the action of the orbifold group.

It is important to note that the branes of type star 
($\star$) in the $(19,19)$ case have different normalization constants 
$N_a$ compared to the non-star branes. Furthermore, for the star 
($\star$) case 
there exists a linear combination of branes with no exceptional part while still being partly rigid. These features provide interesting possibilities for model building.


\subsection{Orientifold Projection}
\label{sec_3_orient_projection}

Since the internal manifold $\M_6$ of this work is compact, one has to cancel the total Ramond-Ramond (R-R) and Neveu Schwarz-Neveu Schwarz (NS-NS) charges of the D6-branes. However, since we are interested in stable and therefore supersymmetric configurations of branes, this can only be achieved by introducing objects with negative tension such as orientifold $6$-planes $[\Pi_{O6}]$. The orientifold planes are the fixed loci of the orientifold projection $\Omega\R(-1)^{F_L}$ where $\Omega$ is the world-sheet parity operator and $\R$ was defined in section \ref{sec_2_orientifold_background}. The symbol $F_L$ denotes the left-moving fermion number and for convenience the factor $(-1)^{F_L}$ will be included in $\Omega$ in the following.

Since the internal manifold is an orbifold, there can be fixed loci not only of $\Omega\R$ but also of $\Omega\R\Theta$, $\Omega\R\Theta'$ and $\Omega\R\Theta\Theta'$. Moreover, there are two different types of contributions to the orientifold plane which will be labelled by $\eta_{\Omega\R g}=\pm 1$. These signs are related to the R-R and NS-NS charge of a brane in the following way
\begin{center}\begin{tabular}{ccccl}
  $\eta_{\Omega\R g}=+1$ &$\qquad$& R-R charge $<0$ &and& NS-NS charge $<0$\,,
  \\
  $\eta_{\Omega\R g}=-1$ &        & R-R charge $>0$ &and& NS-NS charge $>0$\,.
\end{tabular}\end{center}

The condition for cancellation of R-R charges (and therefore also for NS-NS charges) translates into the R-R tadpole cancellation condition. The chiral part of it can be expressed in homology \cite{Aldazabal:2000dg,Blumenhagen:2005mu} as
\begin{equation}\label{tadp_canc}
  \sum_a \N_a\left( [\Pi_a] + [\Pi_a]'\right) = 4\,[\Pi_{O6}]
\end{equation}
where $[\Pi_a]'$ denotes the $\Omega\R$ image of $[\Pi_a]$ and $\N_a$ is the
number of D6-branes wrapping the cycle $[\Pi_a]$.

\bigskip
Let us now turn to the geometric action of $\Omega\R$ on cycles. For the
untwisted $1$-cycles one easily finds that
\begin{equation}\label{orp_fund}\begin{split}
  \Omega\R\,[{a'}^i] &=[{a'}^i]-2\,\beta_i\,[b^i]\,, \\
  \Omega\R\,[b^i] &=-[b^i] \; .
\end{split}\end{equation}
For the bulk-cycle this implies that in $\Omega\R\,[\Pi_a^B]$ one simply has to change the wrapping numbers from $(n_a^i,\tilde{m}_a^i)$ to
$(n_a^i,-\tilde{m}_a^i)$. Such an image of a bulk brane will be denoted as
$\Omega\R\,[\Pi_a^B]=[\widehat{\Pi}^B_a]$ in the following. For the exceptional $2$-cycles one finds \cite{Blumenhagen:2005tn} that $\Omega\,[e^g_{ij}]=-[e^g_{ij}]$. However, also the additional signs $\eta_{\Omega\R g}$ and the permutation of the fixed points under $\R$ have to be taken into account (see table \ref{tab_orp_r}). The final result for the exceptional $3$-cycles then is
\begin{equation}\label{orp_excep}\begin{split}
  \Omega\R\,[\alpha_{ij,n}^g] &=-\eta_{\Omega\R}\,\eta_{\Omega\R g}
    \,[\alpha_{\R(i)\R(j),n}^g]\,, \\
  \Omega\R\,[\alpha_{ij,m}^g] &=+\eta_{\Omega\R}\,\eta_{\Omega\R g}
    \,[\alpha_{\R(i)\R(j),m}^g] \; .
\end{split}\end{equation}
With the help of \eqref{orp_fund} and \eqref{orp_excep} one can compute the orientifold images of the fractional branes $[\Pi_a^{F}]'=\Omega\R\,[\Pi_a^{F}]$. A hat ($\widehat{\hphantom{x}}$) will again indicate the replacement of $\tilde{m}_a^i$ by $-\tilde{m}_a^i$ and in summary one obtains
\begin{align}
  [\Pi_a^{F11}]'=& \hspace{6.05pt}\frac{1}{2}\hspace{6.05pt}\,
    [\widehat{\Pi}^B_a] -
    \eta_{\Omega\R}\,\eta_{\Omega\R\Theta}\hspace{9pt}
    \frac{1}{2}\hspace{9pt}\,\sum_{i,j\in S^{\Theta}_a}
    \epsilon^{\Theta}_{a,ij} [\widehat{\Pi}^{\Theta}_{a,\R(i)\R(j)}]\,, \\
  \begin{split}
  [\Pi_a^{F19}]'=&\frac{1}{N^B_a}\,[\widehat{\Pi}^B_a] - \eta_{\Omega\R}\,
    \eta_{\Omega\R\Theta}\,\frac{1}{N_a^{(1)}}\,
    \sum_{i,j\in S^{\Theta}_a}\epsilon^{\Theta}_{a,ij}
    [\widehat{\Pi}^{\Theta}_{a,\R(i)\R(j)}] \\
  &\hspace{15.8mm}-\eta_{\Omega\R}\,\eta_{\Omega\R\Theta'}\,\frac{1}{N_a^{(2)}}
    \,\sum_{k,l\in S^{\Theta'}_a}\epsilon^{\Theta'}_{a,kl}
    [\widehat{\Pi}^{\Theta'}_{a,\R(k)\R(l)}] \; .
  \end{split}
\end{align}


\subsection{Spectrum and Gauge Groups}
\label{sec_3_gauge_group_spectrum}

The chiral matter content of a given D-brane configuration can be computed in terms of homological intersection numbers of the corresponding cycles in the internal space. For branes not invariant under $\Omega\R$, one finds
gauge groups of the form $\prod_a U(\N_a)$ where $\N_a$ is the number of coincident branes $a$. The chiral fermions transform in bifundamental $(\bar{\Box}_a,\Box_b)$, symmetric $(\Rs_a)$ or antisymmetric ({\scriptsize $\Ratext_{\hspace{-5pt}a}$}) representations of the gauge group. Their multiplicity is determined by the general rules \cite{Blumenhagen:2002wn} displayed in table \ref{tab_rep}. Note that, since the self intersection of $3$-cycles in a $6$-dimensional space is zero, there is no chiral adjoint matter. The explicit form of the needed intersection numbers in the setup of this work is summarized in appendix \ref{app_in}.

\begin{table}[ht]
\begin{center}
\begin{tabular}{>{$}c<{$}||>{$}c<{$}}
  \mbox{Representation} & \mbox{Multiplicity} \\ \hline\hline
  \Ra{a} & \frac{1}{2}\bigl( [\Pi_a]'\circ[\Pi_a] + [\Pi_{O6}]\circ[\Pi_a]
    \bigr) \\
  \Rs_a & \frac{1}{2}\bigl( [\Pi_a]'\circ[\Pi_a] - [\Pi_{O6}]\circ[\Pi_a]
    \bigr) \\
  (\bar{\Box}_a,\Box_b) & [\Pi_a]\circ[\Pi_b] \\
  (\Box_a,\Box_b) & [\Pi_a]'\circ[\Pi_b] \\
  \mbox{Adj}_a & [\Pi_a]\circ [\Pi_a]
\end{tabular}
\end{center}
\caption{\label{tab_rep}Chiral spectrum for intersecting D-branes.}
\end{table}

\bigskip
For cycles which are invariant under the orientifold projection $\Omega\R$ one has to perform a projection on the Chan-Paton degrees of freedom. This will lead to orthogonal or symplectic gauge groups where the precise type has to be determined case by case from the underlying Conformal Field Theory.

However, $\Z_2\times\Z_2$ orientifolds have already been studied in great detail from a CFT point of view. In a type IIB formulation there is the following statement \cite{Bianchi:1991eu,Angelantonj:1999xf,Angelantonj:2002ct} where $g\in\Z_2\times\Z_2$: If a $\Omega g$ invariant D5- or D9-brane wraps the two-torus $\T^2_i$ and there is a non-vanishing constant $B$-field on $\T^2_i$, then one can freely choose $Sp(2\N)$ or $SO(2\N)$ gauge groups for such branes. By a T-dual transformation this result can be translated into the type IIA formulation.
\begin{quote}
If there is a two-torus $\T^2_i$ with $\beta_i=1/2$ and $\Omega\R g$ invariant fractional branes have wrapping numbers $(n_i,\tilde{m}_i)=(\pm 2,0)$ on that $\T^2_i$, then one can freely choose $Sp(2\N)$ {\em or} $SO(2\N)$ gauge groups for such branes.
\end{quote}


\subsection{Consistency and Supersymmetry Conditions}
\label{sec_3_consistency}


\subsubsection*{Tadpole Cancellation}

The R-R tadpole cancellation condition \eqref{tadp_canc} can be made more explicit for the fractional branes constructed in section \ref{sec_3_frac_branes}. It can be split into a bulk part and an exceptional part. For the bulk part one finds
\begin{equation}\label{tadp_bulk}
\begin{array}
{@{}l@{\hspace{3pt}}c@{\hspace{3pt}}l@{\cdot}c@{\cdot}l@{}l@{}l@{}}
 {\displaystyle \sum_a \frac{\N_a}{N^B_a}\: n_a^1\,n_a^2\,n_a^3}
    & = & +4\,\eta_{\Omega\R}
    & 1
    &\Delta_1{\scriptstyle (\lambda_1,0)}
    &\Delta_1{\scriptstyle (\lambda_2,0)}
    &\Delta_1{\scriptstyle (\lambda_3,0)}\;,  \\
 {\displaystyle\sum_a \frac{\N_a}{N^B_a}\:\tilde{m}_a^1\,
    \tilde{m}_a^2\,n_a^3} & =
    & -4\,\eta_{\Omega\R\Theta}
    &2^{-2\beta_1-2\beta_2}
    &\Delta_2{\scriptstyle (\lambda_1,\kappa_1) }
    &\Delta_2{\scriptstyle (\lambda_2,\kappa_2) }
    &\Delta_1{\scriptstyle (\lambda_3,\delta_3) } \;,\\
 {\displaystyle\sum_a \frac{\N_a}{N^B_a}\:n_a^1\,\tilde{m}_a^2\,
    \tilde{m}_a^3} & =
    & -4\,\eta_{\Omega\R\Theta'}
    & 2^{-2\beta_2-2\beta_3}
    &\Delta_1{\scriptstyle (\lambda_1,\delta_1) }
    &\Delta_2{\scriptstyle (\lambda_2,\delta_2+\kappa_2) }
    &\Delta_2{\scriptstyle (\lambda_3,\kappa_3) } \;,\\
 {\displaystyle\sum_a \frac{\N_a}{N^B_a}\:\tilde{m}_a^1\,n_a^2\,
    \tilde{m}_a^3} & =
    &-4\,\eta_{\Omega\R\Theta\Theta'}
    &2^{-2\beta_1-2\beta_3}
    &\Delta_2{\scriptstyle (\lambda_1,\delta_1+\kappa_1) }
    &\Delta_1{\scriptstyle (\lambda_2,\delta_2) }
    &\Delta_2{\scriptstyle (\lambda_3,\delta_3+\kappa_3) }\;,
\end{array}\end{equation}
where $\Delta_1$ and $\Delta_2$ had been defined in equation \eqref{orient_plane_delta}. Here one can see how the various shifts can affect the R-R tadpole cancellation condition. For the exceptional part there is no contribution from the orientifold planes and therefore one calculates
\begin{equation}\begin{split}
  \sum_a \frac{\N_a}{N^{(E_g)}_a}\,n_a^{I_g}\sum_{\{i,j\}\in S_a^g}
    \epsilon^g_{a,ij} \left(
    [\alpha^g_{ij,n}]-\eta_{\Omega\R}\,\eta_{\Omega\R g}\,
    [\alpha^g_{\R(i)\R(j),n}] \right)  &= 0 \,,\\
  \sum_a \frac{\N_a}{N^{(E_g)}_a}\,\tilde{m}_a^{I_g}\sum_{\{i,j\}\in S_a^g}
    \epsilon^g_{a,ij} \left(
    [\alpha^g_{ij,m}]+\eta_{\Omega\R}\,\eta_{\Omega\R g}\,
    [\alpha^g_{\R(i)\R(j),m}] \right)  &= 0\,,
\end{split}\label{tadpole_except}\end{equation}
where $g=\Theta,\Theta'$, $E_g=1,2$ and $I_g=3,1$ for the $\Theta$ and $\Theta'$ sector, respectively.


\subsubsection*{Supersymmetry Conditions}

Naturally, one is interested in stable configurations of D-branes which can be ensured if one considers supersymmetric D-branes preserving the same supersymmetry. This implies that the branes have to satisfy \cite{Becker:1995kb,Blumenhagen:2002wn,Blumenhagen:2005mu}
\begin{equation}\label{slag}
  J |_{\Pi_a} = 0 \qquad\mbox{and}\qquad
  \int_{\Pi_a} \mbox{Im} (e^{i\phi_a}\Omega_3) = 0
\end{equation}
where $\phi_a$ is a phase, $J$ is the K\"ahler-form and $\Omega$ is the holomorphic $3$-form $\Omega_3=dz^1\wedge dz^2\wedge dz^3$. Manifolds with these properties are called Special Lagrangian and they are volume minimizing in their homology class. Therefore in a flat space like $\T^2\times\T^2\times\T^2$, the bulk-branes are straight lines in each $\T^2$ factor.

The phase $\phi_a$ indicates which supersymmetry is preserved. If there are
several branes, one needs them to preserve the same supersymmetry in order to
have a stable configuration. This implies then that the $\phi_a$ have to be
equal for all branes $a$ and orientifold planes. And since $\phi_{O6}=0$ one finds $\phi_a=0$ for all $a$.

\bigskip
The explicit form of \eqref{slag} for the branes in the setup of this work can be found as the following
\begin{equation}\begin{split}
  \tilde{m}^1_a\tilde{m}^2_a\tilde{m}^3_a
    -\frac{n_a^1n_a^2\tilde{m}^3_a}{U^1U^2}
    -\frac{n_a^1\tilde{m}^2_an^3_a}{U^1U^3}
    -\frac{\tilde{m}^1_an^2_an^3_a}{U^2U^3} &= 0\,, \\
  n_a^1n_a^2n_a^3
    -n^1_a\tilde{m}^2_a\tilde{m}^3_aU^2U^3
    -\tilde{m}^1_an^2_a\tilde{m}^3_aU^1U^3
    -\tilde{m}^1_a\tilde{m}^2_an^3_aU^1U^2  &> 0\,,
\end{split}\label{constr_susy}\end{equation}
where $U^i=R^i_y/R^i_x$ are complex structure moduli for $\T^2_i$.


\subsubsection*{K-Theory}

In \cite{Witten:1998cd} it was shown that not all D-brane charges are classified by cohomology but rather by K-theory. Therefore, one has to properly account for charges which may be invisible to ordinary cohomology. In the type of models considered in this work, the additional K-theory charges are usually $\Z_2$ valued and again need to be canceled. However, the computation of K-theory groups for D-branes is very complicated. Instead one can use the field theoretic argument that the cancellation of K-theory charges is equivalent to the absence of global $Sp(2\N)$ anomalies \cite{Witten:1982fp}. This translates into the requirement that the number of $D=4$ chiral fermions transforming in the fundamental representation of some $Sp(2\N)$ factors has to be even.

\bigskip
The global $Sp(2n)$ anomalies can be detected by $Sp(2)$ probe branes
\cite{Uranga:2000xp}. Such probe branes $\Pi_{probe}$ are branes which are
invariant under the orientifold projection $\Omega\R$ and lead to an $Sp(2)$
gauge group. The condition for a model to be free of global $Sp(2n)$ anomalies
is then given by
\begin{equation}\label{ktheory_1}
  [\Pi_{probe}]\circ\sum_a \N_a [\Pi_a^F] \in 2 \Z \; .
\end{equation}
The classification of possible probe branes for a given geometry is very complicated because the invariance under $\Omega\R$ strongly depends on the various shifts $\delta_i$, $\kappa_i$, $\lambda_i$ and on the exceptional charges $\iota,\iota',\iota''$. But equation \eqref{ktheory_1} can be satisfied without knowing about the probe branes if one chooses
\begin{equation}\label{ktheory_2}
  \N_a\in 2\,\Z \qquad \forall a\; .
\end{equation}

\bigskip
Note that the K-theory constraints can also be derived by performing a stability analysis for D-branes as in \cite{Maiden:2006qe}. However, for the models presented in this paper it is sufficient to employ equation \eqref{ktheory_2}.


\subsection{Anomalies and Massive U(1)s}

The anomalies of a given D-brane configuration can be computed in terms of group theoretical quantities which in turn can be expressed as intersection numbers of homological cycles. The potential anomalies for intersecting D-branes are the following and with the help of the tadpole cancellation condition \eqref{tadp_canc} they read as
\begin{equation}\label{anomaly1}\begin{array}{l@{}rcr@{}c}
  \mbox{cubic non-abelian}
    & \mathcal{A}_{SU(\N_a)^3} &=& 0 \hspace{5cm} &\,,\\
  \mbox{mixed abelian}
    & \mathcal{A}_{U(1)_a-SU(\N_b)^2} &=& \frac{1}{2}\,\N_b\,\left( -[\Pi^F_a]+
    [\Pi^F_a]' \right)\circ [\Pi^F_b] &\,, \\
  \mbox{cubic abelian}
    & \mathcal{A}_{U(1)_a-U(1)_b^2} &=& \N_a\,\N_b\,\left( -[\Pi^F_a]+[\Pi^F_a]'
    \right)\circ [\Pi^F_b] &\,, \\
  \mbox{mixed abelian-gravitational}
    & \mathcal{A}_{U(1)_a-G^2} &=& 3\,\N_a\, [\Pi_{O6}]\circ [\Pi^F_a] & \,.
\end{array}\end{equation}

In a consistent theory anomalies have to be canceled. And indeed, string theory provides the Green-Schwarz mechanism \cite{Green:1984sg} to cancel anomalies like \eqref{anomaly1}. For the case of intersecting branes a detailed explanation of the so-called generalized Green-Schwarz mechanism can be found for instance in \cite{Aldazabal:2000dg,Blumenhagen:2005mu}.

However, there are two important consequences of the Green-Schwarz mechanism. First, anomalous $U(1)$ gauge fields will become massive and therefore do not contribute to the chiral spectrum. Secondly, also anomaly free $U(1)$ can become massive and do not show up in the chiral spectrum. The {\em massless} $U(1)$ are given by the kernel of the matrix \cite{Blumenhagen:2005mu}
\begin{equation}\label{anomaly2}
  \N_a\,( [\Pi_{a,I}] - [\Pi_{a,I}]'  )
\end{equation}
where the $3$-cycles $\Pi_a$ have been expanded in an integral basis $\{e^I\}$. Note that the massive $U(1)$ gauge fields do not contribute to the chiral spectrum, but survive as perturbative global symmetries \cite{Ibanez:1999it}.


\clearpage
\section{An Example}
\label{sec_4}

In this section, the methods explained previously will be used to construct a simple example of intersecting D6-branes on a shift $\Z_2\times\Z_2$ orientifold. Although it is relatively easy to find interesting models for the cases with Hodge numbers $(11,11)$ and $(19,19)$, we will choose the $(19,19)$ configuration in order to get completely rigid branes and fix the open-string moduli.


\subsection{Background Geometry and Consistency Conditions}

In general, it is hard to obtain configurations with nice properties like the absence of symmetric and antisymmetric representations in the spectrum. However, one can place D-branes on top of or parallel to orientifold planes. Then the intersection numbers $[\Pi_a]\circ [\Pi_{O6}]$ and $[\Pi_a]'\circ [\Pi_a]$ will vanish trivially and no chiral symmetric or antisymmetric matter will arise. Furthermore, there are no mixed abelian-gravitational anomalies as can be seen from \eqref{anomaly1}.

\bigskip
From the expression for $[\Pi_{O6}]$ \eqref{orient_plane} one finds that there are at most four bulk-branes which are parallel to the orientifold planes. The bulk wrapping numbers $(n^i,\tilde{m}^i)$ of such branes are the following
\begin{equation}
\begin{array}{
  ccr@{\hspace{3pt},\hspace{3pt}}l
     @{\hspace{2pt})\hspace{5pt}\otimes\hspace{5pt}}
    r@{\hspace{3pt},\hspace{3pt}}l
     @{\hspace{2pt})\hspace{5pt}\otimes\hspace{5pt}}
    r@{\hspace{3pt},\hspace{3pt}}l@{\hspace{2pt})\,,}}
  \alpha &:& (\,2^{2\beta_1} \cdot \epsilon_{\alpha}^1 & 0
           & (\,2^{2\beta_2} \cdot \epsilon_{\alpha}^2 & 0
           & (\,2^{2\beta_3} \cdot \epsilon_{\alpha}^3 & 0  \\
  \beta  &:& (\,0 & \epsilon_{\beta}^1
           & (\,0 & \epsilon_{\beta}^2\
           & (\,2^{2\beta_3} \cdot \epsilon_{\beta}^3  & 0  \\
  \gamma &:& (\,2^{2\beta_1} \cdot \epsilon_{\gamma}^1 & 0
           & (\,0 & \epsilon_{\gamma}^2
           & (\,0 & \epsilon_{\gamma}^3                     \\
  \delta &:& (\,0 & \epsilon_{\delta}^1
           & (\,2^{2\beta_2} \cdot \epsilon_{\delta}^2 & 0
           & (\,0 & \epsilon_{\delta}^3
\end{array}
\end{equation}
where $\epsilon_{\xi}^j=\pm 1$ with $\xi=\alpha,\beta,\gamma,\delta$. The
constraints from supersymmetry \eqref{constr_susy} then restrict the $\epsilon_{\xi}^j$ as
\begin{equation}
\begin{array}{r@{\hspace{40pt}}r@{}c}
  \epsilon_{\alpha}^1\cdot\epsilon_{\alpha}^2\cdot\epsilon_{\alpha}^3=+1\,,&
  \epsilon_{\beta}^1\cdot\epsilon_{\beta}^2\cdot\epsilon_{\beta}^3=-1 &\,,\\
  \epsilon_{\gamma}^1\cdot\epsilon_{\gamma}^2\cdot\epsilon_{\gamma}^3=-1\,,&
  \epsilon_{\delta}^1\cdot\epsilon_{\delta}^2\cdot\epsilon_{\delta}^3=-1 &\,.
\end{array}
\label{cons_susy_explicit}
\end{equation}

\bigskip
Let us now choose a background geometry for our example. As it turns out, it is relatively easy to obtain Pati-Salam like models for branes on top of or parallel to the orientifold planes. Therefore we will focus on such configurations.

In order to get suitable ranks for the gauge group factors, we choose one complex structure as $\beta_1=0$ and the others as $\beta_2=\beta_3=1/2$. The consistency condition \eqref{orient_consis_2} then implies that the nontrivial type I shift has to be $\delta_2=\frac{1}{2}\,e^y_2$. The type II and type III shifts are chosen all as $\kappa_i=\lambda_i=0$ in order to get a simple mapping of the fixed points under the orientifold projection $\R$. With this background geometry the R-R tadpole cancellation condition \eqref{tadp_bulk} simplifies as
\begin{equation}\begin{array}{l@{\hspace{60pt}}l}
  {\displaystyle \sum_{\alpha} \frac{\N_{\alpha}}{N^B_{\alpha}} }=
    \eta_{\Omega\R}\;, &
  {\displaystyle \sum_{\beta} \frac{\N_{\beta}}{N^B_{\beta}} }=
    \eta_{\Omega\R\Theta}\;, \\
  {\displaystyle \sum_{\gamma} \frac{\N_{\gamma}}{N^B_{\gamma}} }=
    \eta_{\Omega\R\Theta'}\;, &
  {\displaystyle \sum_{\delta} \frac{\N_{\delta}}{N^B_{\delta}} }=
    \eta_{\Omega\R\Theta\Theta'}\;.
\end{array}\label{cons_tadpole_explicit}
\end{equation}
This in turn implies that all types of branes $\alpha$, $\beta$, $\gamma$ and $\delta$ have to be present and that the charges of the orientifold planes have to be chosen as $\eta_{\Omega\R\,g}=+1$. However, in this setup the charges are not completely unrelated. It can be found for instance in \cite{Blumenhagen:2005tn} that they have to fulfill the condition
\begin{equation}
  \eta = \eta_{\Omega\R}\,\eta_{\Omega\R\Theta}\,\eta_{\Omega\R\Theta'}\,
  \eta_{\Omega\R\Theta\Theta'}
\end{equation}
which implies that the discrete torsion has to be chosen as $\eta=+1$. This is however not a strong restriction for the case of Hodge numbers $(19,19)$ because it fixes only the Wilson line in $\T^2_2$ as one can see from equation \eqref{frac_consistency_2}. In summary, the data to specify the orientifold background for the present example is the following
\begin{equation}\begin{split}
&\begin{array}{lcc@{\hspace{27pt}}lcc@{\hspace{27pt}}lcc@{\hspace{27pt}}lcc}
   \beta_1 &=&0,  &\delta_1 &=&0,         &\kappa_1&=&0,  &\lambda_1&=&0, \\
   \beta_2 &=&1/2,&\delta_2 &=&1/2\,e_2^y,&\kappa_2&=&0,  &\lambda_2&=&0, \\
   \beta_3 &=&1/2,&\delta_3 &=&0,         &\kappa_3&=&0,  &\lambda_3&=&0, \\
\end{array} \\[3mm]
&\begin{array}{lcc@{\hspace{27pt}}lcc@{\hspace{27pt}}lcc}
   \eta_{\Omega\R}             &=&+1,&
   \eta_{\Omega\R\Theta}       &=&+1,&\eta &=&+1, \\
   \eta_{\Omega\R\Theta'}      &=&+1,&
   \eta_{\Omega\R\Theta\Theta'}&=&+1.
\end{array}
\end{split}\end{equation}


\subsection{Fractional Branes}

Since we want to fix the open-string moduli we are interested in rigid branes. This in turn implies that the branes have to be of type star ($\star$). And indeed, comparing the wrapping numbers $(n^2,m^2)=(n^2,\tilde{m}^2-\beta_2\,n^2)$ with table \ref{tab_frac_star} shows that all branes in the present background geometry are completely rigid.

\bigskip
It is easy to see that the bulk intersections of all branes $\alpha$, $\beta$, $\gamma$ and $\delta$ vanish and therefore only the exceptional contributions have to be considered. However, the interesting feature of the $(19,19)$ shift orientifolds is that there are fractional branes which contribute only $1/2$ of a bulk-brane \eqref{half_bulk_brane}. Therefore we choose branes $\beta$ and $\gamma$ as half bulk-branes which results in zero intersection numbers with all others and in the bulk-normalization constant $N^B_{\beta,\gamma}=2$. For brane $\alpha$ and two branes of type $\delta$ we choose pure fractional branes with exceptional sectors which implies $N^B_{\alpha,\delta_1,\delta_2}=4$. For one brane of type $\delta$ we choose the half bulk-brane which gives $N^B_{\delta_3}=2$.

Furthermore, one has to satisfy the consistency condition $\iota'=-\eta$ \eqref{orient_consis_2} which implies that $\iota'_a=-1$ for all branes $a$. These considerations together with a suitable choice of fixed point sets is summarized in table \ref{tab_branes}. Note that the half bulk-branes are not fixed in the second $\T^2$ factor and therefore no fixed points are specified. Since these branes are not completely rigid, not all open string moduli are fixed. This is however no severe problem as we will see in the following.

\begin{table}[ht]
{\footnotesize
\begin{center}\begin{tabular}
  {>{$}c<{$}||
  @{\hspace{3pt}}>{$}c<{$}@{\hspace{3pt}}|
  @{\hspace{3pt}}>{$}c<{$}@{\hspace{3pt}}|
  @{\hspace{3pt}(\hspace{2pt}}>{$}c<{$}@{\hspace{2pt},\hspace{2pt}}>{$}c<{$}
  @{\hspace{2pt})\hspace{2pt}(\hspace{2pt}}>{$}c<{$}
  @{\hspace{2pt},\hspace{2pt}}>{$}c<{$}
  @{\hspace{2pt})\hspace{2pt}(\hspace{2pt}}>{$}c<{$}
  @{\hspace{2pt},\hspace{2pt}}>{$}c<{$}
  @{\hspace{2pt})\hspace{3pt}}|
  @{\hspace{2pt}}>{$}c<{$}@{\hspace{1pt}}>{$}c<{$}@{\hspace{2pt}}|
  @{\hspace{2pt}}>{$}c<{$}@{\hspace{1pt}}>{$}c<{$}@{\hspace{2pt}}|
  @{\hspace{3pt}}>{$}c<{$}@{\hspace{3pt}}>{$}c<{$}
  @{\hspace{3pt}}>{$}c<{$}
  }
  \xi & \N_{\xi} & N^B_{\xi} & n^1 & \tilde{m}^1 & n^2 & \tilde{m}^2 & n^3 &
    \tilde{m}^3 & \{i_1,i_2\} & \{j_1,j_2\} & \{k_1,k_2\} & \{l_1,l_2\} &
    \iota & \iota' & \iota'' \\ \hline\hline
  \alpha  &4&4&+1& 0&+2& 0&+2& 0&\{1,3\}&\{1,2\}&\{3,4\}&\{1,2\}&
    \tau_0&-1&\tau_1 \\
  \beta   &2&2& 0&+1& 0&+1&-2& 0&\{1,2\}&       &       &\{1,2\}&
    \tau_2&-1&\tau_3 \\
  \gamma  &2&2&+1& 0& 0&+1& 0&-1&\{1,3\}&       &       &\{1,2\}&
    \tau_4&-1&\tau_5 \\
  \delta_1&1&4& 0&+1&-2& 0& 0&+1&\{1,2\}&\{1,2\}&\{3,4\}&\{1,2\}&
    \tau_6&-1&\tau_1 \\
  \delta_2&1&4& 0&-1&-2& 0& 0&-1&\{1,2\}&\{1,2\}&\{3,4\}&\{1,2\}&
    \tau_6&-1&\tau_1 \\
  \delta_3&1&2& 0&+1&+2& 0& 0&-1&\{1,2\}&       &       &\{1,2\}&
    \tau_7&-1&\tau_8
\end{tabular}\end{center}
} \caption{\label{tab_branes}Configuration of branes. The $\tau_i$ can be chosen independently as $\pm1$.}
\end{table}

In order to make the structure of the fractional branes more clear, we displayed the branes also in the cycle picture. But one should keep in mind that this way of writing is not complete because not all information about the fixed point sets can be read off.
\begin{equation}\begin{array}
{@{}l@{\hspace{5pt}}c@{\hspace{5pt}}r@{\hspace{3pt}}c@{}c@{}c@{}c@{}c
@{\hspace{3pt}}r@{\hspace{3pt}}l@{\hspace{3pt}}r@{\hspace{3pt}}l@{}}
  [\alpha]  &=&+           &[a^1]&\otimes&[a^2]&\otimes&[a^3]
    &+           &\left([\alpha^{\Theta}_{11,n}]
                       +\tau_0[\alpha^{\Theta}_{31,n}]\right)
    &+\frac{1}{2}&\left([\alpha^{\Theta'}_{31,n}]
                       +\tau_1[\alpha^{\Theta'}_{32,n}]\right) \\[1mm]
  [\beta]   &=&-           &[b^1]&\otimes&[b^2]&\otimes&[a^3]  \\[1mm]
  [\gamma]  &=&-\frac{1}{2}&[a^1]&\otimes&[b^2]&\otimes&[b^3]  \\[1mm]
  [\delta_1]&=&-\frac{1}{2}&[b^1]&\otimes&[a^2]&\otimes&[b^3]
    &+\frac{1}{2}&\left([\alpha^{\Theta}_{11,m}]
                       +\tau_6[\alpha^{\Theta}_{21,m}]\right)
    &+\frac{1}{2}&\left([\alpha^{\Theta'}_{31,m}]
                       +\tau_1[\alpha^{\Theta'}_{32,m}]\right) \\[1mm]
  [\delta_2]&=&-\frac{1}{2}&[b^1]&\otimes&[a^2]&\otimes&[b^3]
    &-\frac{1}{2}&\left([\alpha^{\Theta}_{11,m}]
                       +\tau_6[\alpha^{\Theta}_{21,m}]\right)
    &-\frac{1}{2}&\left([\alpha^{\Theta'}_{31,m}]
                       +\tau_1[\alpha^{\Theta'}_{32,m}]\right) \\[1mm]
  [\delta_3]&=&-           &[b^1]&\otimes&[a^2]&\otimes&[b^3]&
\end{array}\end{equation}

\bigskip
One can check that the configuration of branes in table \ref{tab_branes} satisfies the R-R tadpole cancellation condition \eqref{cons_tadpole_explicit} and \eqref{tadpole_except} and the supersymmetry conditions \eqref{cons_susy_explicit}. For the K-theory constraints note that except for the $\delta$ branes there is always an even number of branes $\N$. For the last three branes one can see that $\delta_1+\delta_2$ add up to half a bulk-brane in a homological sense and have therefore zero intersections with the possible probe branes. And since $\delta_3$ has no exceptional part, the same applies also to this brane. Therefore, all consistency conditions for the configuration of branes in table \ref{tab_branes} are fulfilled.


\subsection{Gauge Groups and Spectrum}

To determine the type of gauge group one first has to check which branes are invariant under the orientifold projection $\Omega\R$. As it turns out, only brane $\alpha$ is not invariant and therefore leads to the gauge group $U(4)$. The other branes are invariant under $\Omega\R$ and therefore their gauge group is $SO(2\N)$ or $Sp(2\N)$. From the explanation at the end of section \ref{sec_3_gauge_group_spectrum} it follows that for $\beta$ and $\delta$ one can choose symplectic gauge groups because the wrapping numbers satisfy $(n_3,\tilde{m}_3)=(-2,0)$ and $(n_2,\tilde{m}_2)=(\mp2,0)$, respectively. For brane $\gamma$ the precise type of gauge group has to be determined from the underlying Conformal Field Theory.

\bigskip
With the help of table \ref{tab_rep} one can compute the chiral spectrum from the homological intersection numbers between the various cycles. The result is displayed in table \ref{tab_spec} where it was used that the $\bar{2}$ representation of $Sp(2)$ is isomorphic to the $2$ representation. Furthermore, from equation \eqref{anomaly1} one finds that the $U(1)$ factor of brane $\alpha$ is anomalous. Therefore it receives a mass by the Generalized Green-Schwarz mechanism and does not contribute to the chiral spectrum.

\begin{table}[ht]
{\footnotesize
\begin{center}\begin{tabular}
  {>{$}c<{$}||>{$}r<{$}@{}>{$}c<{$}@{}>{$}c<{$}@{}>{$}c<{$}
  @{\hspace{1mm}}>{$}c<{$}@{\hspace{1mm}}>{$}c<{$}@{}>{$}c<{$}
  @{}>{$}c<{$}@{}>{$}c<{$}}
  \mbox{branes} & & U(4)_{\alpha}\times& Sp(2)_{\delta_1}\times
    &Sp(2)_{\delta_2}&\times&Sp(4)_{\beta}\times
    &\frac{Sp(4)}{SO(4)}_{\gamma}\times&Sp(2)_{\delta_3} \\
    \hline\hline
  \alpha \delta_1&2\times(&\bar{4},&2,&1,&&1,&1,&1&) \\
  \alpha'\delta_1&2\times(&\bar{4},&2,&1,&&1,&1,&1&)\\
  \alpha \delta_2&2\times(&4,&1,&2,&&1,&1,&1&) \\
  \alpha'\delta_2&2\times(&4,&1,&2,&&1,&1,&1&) \\
\end{tabular}\end{center}
} \caption{\label{tab_spec}Chiral spectrum resulting from the branes of table \ref{tab_branes}.}
\end{table}

\bigskip
From table \ref{tab_spec} one can see that there are no chiral representations transforming under the gauge factor $Sp(4)_{\beta} \times \frac{Sp(4)}{SO(4)}_{\gamma} \times Sp(2)_{\delta_3}$. Therefore this part can be considered as to be in the hidden sector. Moreover, since the unfixed open-string moduli of this example come from the branes $\beta$, $\gamma$ and $\delta_3$ they are not relevant. For the visible sector the total gauge group is
\begin{equation}
  SU(4)_{\alpha} \times SU(2)_{\delta_1} \times
  SU(2)_{\delta_2}
\end{equation}
where $Sp(2)\cong SU(2)$ has been employed. Since the branes $\delta$ are invariant under the orientifold projection, the representations resulting from $\alpha\circ\delta_{1,2}$ and $\alpha'\circ\delta_{1,2}$ are actually the same. Therefore the chiral matter content of this example is
\begin{equation}\begin{split}
  &2 \times (\overline{4},2,1)_{-1} \\
  &2 \times (4,1,2)_{+1}
\end{split}\end{equation}
where the subindex indicates the $U(1)_{\alpha}$ charge. Note that this is a two generation Pati-Salam like model where all visible branes are completely rigid and no open-string moduli fields appear for the visible sector.


\section{Summary and Outlook}
\label{sec_5}

In this paper we have presented a complete geometrical analysis of shift
$\Z_2\times\Z_2$ orientifolds from an intersecting brane perspective. As it turned out, the type I shifts $\delta$ determine the topology of the orbifold and therefore the number of homological cycles and R-R tadpole cancellation conditions. The type II and type III shifts $\kappa$ and $\lambda$ are responsible for the permutation of fixed points under the orientifold projection $\Omega\R$ and they strongly influence the
presence of the orientifold planes. In particular, depending on $\kappa$ and
$\lambda$ it is possible to have zero, one two, three or four sectors
contributing to the orientifold plane. This in turn implies, that the R-R
tadpole cancellation condition can be modified by these shifts.

In the second part of this work we explicitly constructed fractional $3$-cycles
on the orbifold and developed the necessary techniques to analyze intersecting
D6-brane models. Using these methods, in section \ref{sec_4} a simple but
interesting model was presented. This example is of type Pati-Salam with two
generations and {\em no} open-string moduli (in the visible sector). This is clearly not a realistic model, but since its construction is very simple one might hope to find more realistic models in more general setups. Eventually one would also like to turn on additional fluxes and search for realistic models
in this extended set-up \cite{btl03,cu03,Marchesano:2004yq,Cvetic:2004xx,ms04a,Camara:2005dc}.
However, since there is a huge number of different orientifold geometries one might need to perform a computer search to look for interesting configurations or even perform a landscape study like in \cite{Dijkstra:2004cc,Blumenhagen:2004xx,Gmeiner:2005vz}.

Another direction of study would be to consider other orbifold groups like
$\Z_3$. For the corresponding orientifold one can show that non-trivial shifts will again modify the presence of orientifold planes and hence the R-R tadpole cancellation condition.


\subsection*{Acknowledgments}

We would like to thank  M. Cveti{\v c} and G. Shiu for collaboration
and discussions at early stages of this work. E.P. is grateful for helpful
discussions with G. Honecker and T. Weigand and wants to thank the Max-Planck-Institute for Physics, in particular 
D. L\"ust for support.


\cleardoublepage
\begin{appendix}


\clearpage
\section{Summary of Intersection Numbers}
\label{app_in}

To compute the chiral spectrum of table \ref{tab_rep}, one needs the
intersection numbers between fractional branes $[\Pi^F_a]$, their orientifold
images $[\Pi^F_a]'$ and the orientifold planes $[\Pi_{O6}]$. In the following
these combinations are summarized. For convenience, the notation $I_{ab}^i=n_a^i\tilde{m}_b^i-\tilde{m}_a^in_b^i$ and $\widehat{I}_{ab}^i=n_a^i\tilde{m}_b^i+\tilde{m}_a^in_b^i$ and the definitions in \eqref{def_deltas} will be used
\begin{equation}\label{def_deltas}\begin{array}{lcl@{}l@{}l}
  \Delta^{(11)}_{1}            &=&         \Delta_1(\lambda_1,0)
                                 &\:\cdot\:\Delta_1(\lambda_2,0)
                                 &\:\cdot\:\Delta_1(\lambda_3,0) \;, \\
  \Delta^{(11)}_{\Theta}       &=&         \Delta_2(\lambda_1,\kappa_1)
                                 &\:\cdot\:\Delta_2(\lambda_2,\kappa_2)
                                 &\:\cdot\:\Delta_1(\lambda_3,0) \;, \\
  \Delta^{(11)}_{\Theta'}      &=&         \Delta_1(\lambda_1,\delta_1)
                                 &\:\cdot\:\Delta_2(\lambda_2,\delta_2+\kappa_2)
                                 &\:\cdot\:\Delta_2(\lambda_3\kappa_3)\;, \\
  \Delta^{(11)}_{\Theta\Theta'}&=&         \Delta_2(\lambda_1,\delta_1+\kappa_1)
                                 &\:\cdot\:\Delta_1(\lambda_2,\delta_2)
                                 &\:\cdot\:\Delta_2(\lambda_3,\kappa_3)\,,
                                 \\[10pt]
  \Delta^{(19)}_{1}            &=&         \Delta_1(\lambda_1,0)
                                 &\:\cdot\:\Delta_1(\lambda_2,0)
                                 &\:\cdot\:\Delta_1(\lambda_3,0) \;, \\
  \Delta^{(19)}_{\Theta}       &=&         \Delta_2(\lambda_1,\kappa_1)
                                 &\:\cdot\:\Delta_2(\lambda_2,\kappa_2)
                                 &\:\cdot\:\Delta_1(\lambda_3,0) \;, \\
  \Delta^{(19)}_{\Theta'}      &=&         \Delta_1(\lambda_1,0)
                                 &\:\cdot\:\Delta_2(\lambda_2,\delta_2+\kappa_2)
                                 &\:\cdot\:\Delta_2(\lambda_3\kappa_3)\;, \\
  \Delta^{(19)}_{\Theta\Theta'}&=&         \Delta_2(\lambda_1,\kappa_1)
                                 &\:\cdot\:\Delta_1(\lambda_2,\delta_2)
                                 &\:\cdot\:\Delta_2(\lambda_3,\kappa_3)\,.
\end{array}\end{equation}


\subsection{The Case with Hodge Numbers (11,11)}

\begin{align}
  \begin{split}
  [\Pi^{F11}_a]\circ[\Pi^{F11}_b]=& I_{ab}
    +\frac{1}{2}\, I_{ab}^3
    \sum_{\genfrac{}{}{0pt}{}{\{i,j\}\in S^{\Theta}_a}{\{s,t\}
    \in S^{\Theta}_b}}
    \epsilon^{\Theta}_{a,ij}\epsilon^{\Theta}_{b,st} \left( \delta_{is}
    \delta_{jt}-\eta\, \delta_{ip(s)}\delta_{jq(t)} \right)
  \end{split} \\
  \begin{split}
  [\Pi^{F11}_a]'\circ[\Pi^{F11}_b]=&\widehat{I}_{ab}
    -\eta_{\Omega\R}\,\eta_{\Omega\R\Theta}\,\frac{1}{2}\,
    \widehat{I}_{ab}^3 \\
    &\hphantom{\frac{4}{N^B_aN^B_b}\, \widehat{I}_{ab}-}
    \cdot\sum_{\genfrac{}{}{0pt}{}{\{i,j\}\in S^{\Theta}_a}{\{s,t\}
    \in S^{\Theta}_b}}
    \epsilon^{\Theta}_{a,ij}\epsilon^{\Theta}_{b,st} \left( \delta_{\R(i)s}
    \delta_{\R(j)t}-\eta\, \delta_{\R(i)p(s)}\delta_{\R(j)q(t)} \right)
  \end{split} \\
  \begin{split}
  [\Pi^{F11}_a]'\circ[\Pi^{F11}_a]=&8\,\prod_{i=1}^3
    \left(n_a^i\tilde{m}_a^i\right)
    -\eta_{\Omega\R}\,\eta_{\Omega\R\Theta}\,
    \left(n_a^3\tilde{m}_a^3\right) \\
    &\hphantom{\frac{4}{N^B_aN^B_b}\, \widehat{I}_{ab}-}
    \cdot\sum_{\genfrac{}{}{0pt}{}{\{i,j\}\in S^{\Theta}_a}{\{s,t\}
    \in S^{\Theta}_a}}
    \epsilon^{\Theta}_{a,ij}\epsilon^{\Theta}_{a,st} \left( \delta_{\R(i)s}
    \delta_{\R(j)t}-\eta\, \delta_{\R(i)p(s)}\delta_{\R(j)q(t)} \right)
  \end{split} \\
  \begin{split}
  [\Pi_{O6}]\circ[\Pi_a^{F11}]=&4\,\bigl(
  \begin{array}[t]{r@{\hspace{1.5mm}}c@{\hspace{1mm}}c@{\hspace{1mm}}
                    c@{\hspace{1.5mm}}l@{\hspace{1.5mm}}l}
    \eta_{\Omega\R}&\tilde{m}_a^1&\tilde{m}_a^2&\tilde{m}_a^3&
      &\Delta^{(11)}_{1}  \\
    -\eta_{\Omega\R\Theta}&n_a^1&n_a^2&\tilde{m}_a^3&
      2^{-2\beta_1-2\beta_2}&\Delta^{(11)}_{\Theta}  \\
    -\eta_{\Omega\R\Theta'}&\tilde{m}_a^1&n_a^2&n_a^3&
      2^{-2\beta_2-2\beta_3}&\Delta^{(11)}_{\Theta'}\\
    -\eta_{\Omega\R\Theta\Theta'}&n_a^1&\tilde{m}_a^2&n_a^3&
      2^{-2\beta_1-2\beta_3}&\Delta^{(11)}_{\Theta\Theta'} \;\bigl.\bigr)
  \end{array} \bigr.
  \end{split}
\end{align}


\subsection{The Case with Hodge Numbers (19,19)}

\begin{align}
  \begin{split}
  [\Pi_a^{F19}]\circ[\Pi_b^{F19}]=&\hphantom{+}\frac{4}{N^B_aN^B_b}\,I_{ab}
    +\frac{2}{N^{(1)}_aN^{(1)}_b}\,I_{ab}^3
    \sum_{\genfrac{}{}{0pt}{}{\{i,j\}\in S^{\Theta}_a}{\{s,t\}\in
    S^{\Theta}_b}}
    \epsilon^{\Theta}_{a,ij}\epsilon^{\Theta}_{b,st}\,\delta_{is}\left(
    \delta_{jt}-\eta\,\delta_{jq(t)}\right) \\
  &+\frac{2}{N^{(2)}_aN^{(2)}_b}\,I_{ab}^1
    \sum_{\genfrac{}{}{0pt}{}{\{k,l\}\in S^{\Theta'}_a}{\{u,v\}\in
    S^{\Theta'}_b}}
    \epsilon^{\Theta'}_{a,kl}\epsilon^{\Theta'}_{b,uv}\left(
    \delta_{ku}-\eta\,\delta_{kq(u)}\right)\,\delta_{lv}
  \end{split} \\
  \begin{split}
  [\Pi_a^{F19}]'\circ[\Pi_b^{F19}]=&\hphantom{+}\frac{4}{N^B_aN^B_b}\,
    \widehat{I}_{ab}-\eta_{\Omega\R}\,\eta_{\Omega\R\Theta}\,
    \frac{2}{N^{(1)}_aN^{(1)}_b}\,\widehat{I}_{ab}^3 \\
  &\cdot\sum_{\genfrac{}{}{0pt}{}{\{i,j\}\in S^{\Theta}_a}{\{s,t\}\in
    S^{\Theta}_b}}
    \epsilon^{\Theta}_{a,ij}\epsilon^{\Theta}_{b,st}\,\delta_{\R
    (i)s}\left(
    \delta_{\R(j)t}-\eta\,\delta_{\R(j)q(t)}\right)
    -\eta_{\Omega\R}\,\eta_{\Omega\R\Theta'} \\
  &\cdot\frac{2}{N^{(2)}_aN^{(2)}_b}\,\widehat{I}_{ab}^1
    \sum_{\genfrac{}{}{0pt}{}{\{k,l\}\in S^{\Theta'}_a}{\{u,v\}\in
    S^{\Theta'}_b}}
    \epsilon^{\Theta'}_{a,kl}\epsilon^{\Theta'}_{b,uv}\left(
    \delta_{\R(k)u}-\eta\,\delta_{\R(k)q(u)}\right)\,\delta_{\R(l)v}
  \end{split} \\
  \begin{split}
  [\Pi_a^{F19}]'\circ[\Pi_a^{F19}]=&\hphantom{+}\frac{32}{N^{B\, 2}_a}\,
    \prod_{i=1}^3\left(n_a^i\tilde{m}_a^i\right)
    -\eta_{\Omega\R}\,\eta_{\Omega\R\Theta}\,\frac{4}{N^{(1)\, 2}_a}\,
    \left(n_a^3\tilde{m}_a^3\right) \\
  &\cdot\sum_{\genfrac{}{}{0pt}{}{\{i,j\}\in S^{\Theta}_a}{s,t\in
    S^{\Theta}_a}}
    \epsilon^{\Theta}_{a,ij}\epsilon^{\Theta}_{a,st}\,\delta_{\R
    (i)s}\left(\delta_{\R(j)t}-\eta\,\delta_{\R
    (j)q(t)}\right)-\eta_{\Omega\R}\,\eta_{\Omega\R\Theta'} \\
  &\cdot\frac{4}{N^{(2)\, 2}_a}\,\left(n_a^1\tilde{m}_a^1\right)
    \sum_{\genfrac{}{}{0pt}{}{\{k,l\}\in S^{\Theta'}_a}{\{u,v\}\in
    S^{\Theta'}_b}}
    \epsilon^{\Theta'}_{a,kl}\epsilon^{\Theta'}_{b,uv}\left(
    \delta_{\R(k)u}-\eta\,\delta_{\R(k)q(u)}\right)\,\delta_{\R(l)v}
  \end{split} \\
  \begin{split}
  [\Pi_{O6}]\circ[\Pi_a^{F19}]=&\hphantom{+}\frac{8}{N^B_a}\bigl(
  \begin{array}[t]{r@{\hspace{1.5mm}}c@{\hspace{1mm}}c@{\hspace{1mm}}
                    c@{\hspace{1.5mm}}l@{\hspace{1.5mm}}l}
    \eta_{\Omega\R}&\tilde{m}_a^1&\tilde{m}_a^2&\tilde{m}_a^3&
      &\Delta^{(19)}_{1}  \\
    -\eta_{\Omega\R\Theta}&n_a^1&n_a^2&\tilde{m}_a^3&
      2^{-2\beta_1-2\beta_2}&\Delta^{(19)}_{\Theta}  \\
    -\eta_{\Omega\R\Theta'}&\tilde{m}_a^1&n_a^2&n_a^3&
      2^{-2\beta_2-2\beta_3}&\Delta^{(19)}_{\Theta'} \\
    -\eta_{\Omega\R\Theta\Theta'}&n_a^1&\tilde{m}_a^2&n_a^3&
      2^{-2\beta_1-2\beta_3}&\Delta^{(19)}_{\Theta\Theta'} \;\bigl.\bigr)
  \end{array} \bigr.
  \end{split}
\end{align}


\end{appendix}


\cleardoublepage \nocite{*}
\bibliography{references}
\bibliographystyle{utphys}


\end{document}

%% file: fixpts_10
\setlength{\unitlength}{3947sp}%
\begingroup\makeatletter\ifx\SetFigFont\undefined%
\gdef\SetFigFont#1#2#3#4#5{%
  \reset@font\fontsize{#1}{#2pt}%
  \fontfamily{#3}\fontseries{#4}\fontshape{#5}%
  \selectfont}%
\fi\endgroup%
\begin{picture}(1009,1009)(1104,-1658)
{\color[rgb]{0,0,0}\thinlines
\put(1201,-1561){\circle*{72}}
}%
{\color[rgb]{0,0,0}\put(1651,-1561){\circle*{72}}
}%
{\color[rgb]{0,0,0}\put(1201,-1111){\circle*{72}}
}%
{\color[rgb]{0,0,0}\put(1651,-1111){\circle*{72}}
}%
{\color[rgb]{0,0,0}\put(1201,-1561){\framebox(900,900){}}
}%
\end{picture}%

%% file: fixpts_11
\setlength{\unitlength}{3947sp}%
\begingroup\makeatletter\ifx\SetFigFont\undefined%
\gdef\SetFigFont#1#2#3#4#5{%
  \reset@font\fontsize{#1}{#2pt}%
  \fontfamily{#3}\fontseries{#4}\fontshape{#5}%
  \selectfont}%
\fi\endgroup%
\begin{picture}(1009,1009)(1104,-1658)
{\color[rgb]{0,0,0}\thinlines
\put(1876,-886){\circle*{72}}
}%
{\color[rgb]{0,0,0}\put(1876,-1336){\circle*{72}}
}%
{\color[rgb]{0,0,0}\put(1426,-1336){\circle*{72}}
}%
{\color[rgb]{0,0,0}\put(1426,-886){\circle*{72}}
}%
{\color[rgb]{0,0,0}\put(1201,-1561){\framebox(900,900){}}
}%
\end{picture}%

%% file: fixpts_none
\setlength{\unitlength}{3947sp}%
\begingroup\makeatletter\ifx\SetFigFont\undefined%
\gdef\SetFigFont#1#2#3#4#5{%
  \reset@font\fontsize{#1}{#2pt}%
  \fontfamily{#3}\fontseries{#4}\fontshape{#5}%
  \selectfont}%
\fi\endgroup%
\begin{picture}(1009,1009)(1104,-1658)
\thinlines
{\color[rgb]{0,0,0}\put(1201,-1561){\framebox(900,900){}}
}%
\end{picture}%

%% file: figure_01
\setlength{\unitlength}{3947sp}

\begin{picture}(5300,2150)(49,900)

\thinlines

\put(750,2100){\circle*{70}}
\put(750,1500){\circle*{70}}
\put(1350,2100){\circle*{70}}
\put(1350,1500){\circle*{70}}

\put(3750,1650){\circle*{70}}
\put(3750,2250){\circle*{70}}
\put(4350,2550){\circle*{70}}
\put(4350,1950){\circle*{70}}

\put(450,1200){\line( 1, 0){1800}}
\put(375,2400){\line( 1, 0){1275}}
\put(450,1200){\line( 0, 1){1800}}
\put(1650,2400){\line( 0,-1){1275}}

\put(3450,1200){\line( 1, 0){1800}}
\put(3450,1200){\line( 0, 1){1800}}
\put(3375,2400){\line( 1, 0){150}}
\put(4650,1275){\line( 0,-1){150}}
\put(4651,3000){\line( 0,-1){1200}}
\put(3450,2400){\line( 2, 1){1200}}
\put(3450,1200){\line( 2, 1){1200}}

\thicklines

\put(450,1200){\vector( 1, 0){1200}}
\put(450,1200){\vector( 0, 1){1200}}
\put(3450,1200){\vector( 2, 1){1200}}
\put(3450,1200){\vector( 0, 1){1200}}

\put(900,900){\makebox{$[a^i]$}}
\put(80,1750){\makebox{$[b^i]$}}
\put(3080,1750){\makebox{$[b^i]$}}
\put(4600,1500){\makebox{$[{a'}^i]$}}

\put(1460,900){\makebox{\footnotesize $2\pi R^x_i$}}
\put(4460,900){\makebox{\footnotesize $2\pi R^x_i$}}
\put(-150,2340){\makebox{\footnotesize $2\pi i\,R^y_i$}}
\put(2850,2350){\makebox{\footnotesize $2\pi i\,R^y_i$}}

\put(600,1400){\makebox{\footnotesize 1}}
\put(600,2000){\makebox{\footnotesize 2}}
\put(1200,1400){\makebox{\footnotesize 3}}
\put(1200,2000){\makebox{\footnotesize 4}}
\put(3600,1550){\makebox{\footnotesize 1}}
\put(3600,2150){\makebox{\footnotesize 2}}
\put(4200,1850){\makebox{\footnotesize 3}}
\put(4200,2450){\makebox{\footnotesize 4}}

\put(1800,2700){\makebox{$\beta_i=0$}}
\put(4800,2700){\makebox{$\beta_i=\frac{1}{2}$}}

\end{picture}

%% file: fixpts_31
\setlength{\unitlength}{3947sp}%
\begingroup\makeatletter\ifx\SetFigFont\undefined%
\gdef\SetFigFont#1#2#3#4#5{%
  \reset@font\fontsize{#1}{#2pt}%
  \fontfamily{#3}\fontseries{#4}\fontshape{#5}%
  \selectfont}%
\fi\endgroup%
\begin{picture}(1009,1009)(1104,-1658)
\thinlines
{\color[rgb]{0,0,0}\put(1201,-1561){\framebox(900,900){}}
}%
{\color[rgb]{0,0,0}\put(1801,-961){\framebox(150,150){}}
}%
{\color[rgb]{0,0,0}\put(1351,-961){\framebox(150,150){}}
}%
{\color[rgb]{0,0,0}\put(1351,-1411){\framebox(150,150){}}
}%
{\color[rgb]{0,0,0}\put(1801,-1411){\framebox(150,150){}}
}%
\end{picture}%

%% file: fixpts_21
\setlength{\unitlength}{3947sp}%
\begingroup\makeatletter\ifx\SetFigFont\undefined%
\gdef\SetFigFont#1#2#3#4#5{%
  \reset@font\fontsize{#1}{#2pt}%
  \fontfamily{#3}\fontseries{#4}\fontshape{#5}%
  \selectfont}%
\fi\endgroup%
\begin{picture}(1009,1009)(1104,-1658)
\thinlines
{\color[rgb]{0,0,0}\put(1201,-1561){\framebox(900,900){}}
}%
\thicklines
{\color[rgb]{0,0,0}\multiput(1351,-1411)(6.00000,6.00000){26}{\makebox(6.6667,10.0000){\SetFigFont{7}{8.4}{\rmdefault}{\mddefault}{\updefault}.}}
}%
{\color[rgb]{0,0,0}\multiput(1351,-1261)(6.00000,-6.00000){26}{\makebox(6.6667,10.0000){\SetFigFont{7}{8.4}{\rmdefault}{\mddefault}{\updefault}.}}
}%
{\color[rgb]{0,0,0}\multiput(1801,-1411)(6.00000,6.00000){26}{\makebox(6.6667,10.0000){\SetFigFont{7}{8.4}{\rmdefault}{\mddefault}{\updefault}.}}
}%
{\color[rgb]{0,0,0}\multiput(1801,-1261)(6.00000,-6.00000){26}{\makebox(6.6667,10.0000){\SetFigFont{7}{8.4}{\rmdefault}{\mddefault}{\updefault}.}}
}%
{\color[rgb]{0,0,0}\multiput(1801,-811)(6.00000,-6.00000){26}{\makebox(6.6667,10.0000){\SetFigFont{7}{8.4}{\rmdefault}{\mddefault}{\updefault}.}}
}%
{\color[rgb]{0,0,0}\multiput(1801,-961)(6.00000,6.00000){26}{\makebox(6.6667,10.0000){\SetFigFont{7}{8.4}{\rmdefault}{\mddefault}{\updefault}.}}
}%
{\color[rgb]{0,0,0}\multiput(1351,-811)(6.00000,-6.00000){26}{\makebox(6.6667,10.0000){\SetFigFont{7}{8.4}{\rmdefault}{\mddefault}{\updefault}.}}
}%
{\color[rgb]{0,0,0}\multiput(1351,-961)(6.00000,6.00000){26}{\makebox(6.6667,10.0000){\SetFigFont{7}{8.4}{\rmdefault}{\mddefault}{\updefault}.}}
}%
\end{picture}%

%% file: fixpts_20
\setlength{\unitlength}{3947sp}%
\begingroup\makeatletter\ifx\SetFigFont\undefined%
\gdef\SetFigFont#1#2#3#4#5{%
  \reset@font\fontsize{#1}{#2pt}%
  \fontfamily{#3}\fontseries{#4}\fontshape{#5}%
  \selectfont}%
\fi\endgroup%
\begin{picture}(1009,1009)(1104,-1658)
\thinlines
{\color[rgb]{0,0,0}\put(1201,-1561){\framebox(900,900){}}
}%
\thicklines
{\color[rgb]{0,0,0}\multiput(1126,-1186)(6.00000,6.00000){26}{\makebox(6.6667,10.0000){\SetFigFont{7}{8.4}{\rmdefault}{\mddefault}{\updefault}.}}
}%
{\color[rgb]{0,0,0}\multiput(1126,-1636)(6.00000,6.00000){26}{\makebox(6.6667,10.0000){\SetFigFont{7}{8.4}{\rmdefault}{\mddefault}{\updefault}.}}
}%
{\color[rgb]{0,0,0}\multiput(1576,-1636)(6.00000,6.00000){26}{\makebox(6.6667,10.0000){\SetFigFont{7}{8.4}{\rmdefault}{\mddefault}{\updefault}.}}
}%
{\color[rgb]{0,0,0}\multiput(1576,-1186)(6.00000,6.00000){26}{\makebox(6.6667,10.0000){\SetFigFont{7}{8.4}{\rmdefault}{\mddefault}{\updefault}.}}
}%
{\color[rgb]{0,0,0}\multiput(1576,-1036)(6.00000,-6.00000){26}{\makebox(6.6667,10.0000){\SetFigFont{7}{8.4}{\rmdefault}{\mddefault}{\updefault}.}}
}%
{\color[rgb]{0,0,0}\multiput(1126,-1036)(6.00000,-6.00000){26}{\makebox(6.6667,10.0000){\SetFigFont{7}{8.4}{\rmdefault}{\mddefault}{\updefault}.}}
}%
{\color[rgb]{0,0,0}\multiput(1126,-1486)(6.00000,-6.00000){26}{\makebox(6.6667,10.0000){\SetFigFont{7}{8.4}{\rmdefault}{\mddefault}{\updefault}.}}
}%
{\color[rgb]{0,0,0}\multiput(1576,-1486)(6.00000,-6.00000){26}{\makebox(6.6667,10.0000){\SetFigFont{7}{8.4}{\rmdefault}{\mddefault}{\updefault}.}}
}%
\end{picture}%

%% file: fixpts_11_30
\setlength{\unitlength}{3947sp}%
\begingroup\makeatletter\ifx\SetFigFont\undefined%
\gdef\SetFigFont#1#2#3#4#5{%
  \reset@font\fontsize{#1}{#2pt}%
  \fontfamily{#3}\fontseries{#4}\fontshape{#5}%
  \selectfont}%
\fi\endgroup%
\begin{picture}(1009,1009)(1104,-1658)
{\color[rgb]{0,0,0}\thinlines
\put(1201,-1561){\circle*{72}}
}%
{\color[rgb]{0,0,0}\put(1651,-1561){\circle*{72}}
}%
{\color[rgb]{0,0,0}\put(1651,-1111){\circle*{72}}
}%
{\color[rgb]{0,0,0}\put(1201,-1111){\circle*{72}}
}%
{\color[rgb]{0,0,0}\put(1201,-1561){\framebox(900,900){}}
}%
{\color[rgb]{0,0,0}\put(1801,-961){\framebox(150,150){}}
}%
{\color[rgb]{0,0,0}\put(1801,-1411){\framebox(150,150){}}
}%
{\color[rgb]{0,0,0}\put(1351,-1411){\framebox(150,150){}}
}%
{\color[rgb]{0,0,0}\put(1351,-961){\framebox(150,150){}}
}%
\end{picture}%

%% file: fixpts_30
\setlength{\unitlength}{3947sp}%
\begingroup\makeatletter\ifx\SetFigFont\undefined%
\gdef\SetFigFont#1#2#3#4#5{%
  \reset@font\fontsize{#1}{#2pt}%
  \fontfamily{#3}\fontseries{#4}\fontshape{#5}%
  \selectfont}%
\fi\endgroup%
\begin{picture}(1009,1009)(1104,-1658)
\thinlines
{\color[rgb]{0,0,0}\put(1201,-1561){\framebox(900,900){}}
}%
{\color[rgb]{0,0,0}\put(1126,-1636){\framebox(150,150){}}
}%
{\color[rgb]{0,0,0}\put(1576,-1636){\framebox(150,150){}}
}%
{\color[rgb]{0,0,0}\put(1576,-1186){\framebox(150,150){}}
}%
{\color[rgb]{0,0,0}\put(1126,-1186){\framebox(150,150){}}
}%
\end{picture}%

%% file: fixpts_10_21
\setlength{\unitlength}{3947sp}%
\begingroup\makeatletter\ifx\SetFigFont\undefined%
\gdef\SetFigFont#1#2#3#4#5{%
  \reset@font\fontsize{#1}{#2pt}%
  \fontfamily{#3}\fontseries{#4}\fontshape{#5}%
  \selectfont}%
\fi\endgroup%
\begin{picture}(1009,1009)(1104,-1658)
{\color[rgb]{0,0,0}\thinlines
\put(1651,-1111){\circle*{72}}
}%
{\color[rgb]{0,0,0}\put(1651,-1561){\circle*{72}}
}%
{\color[rgb]{0,0,0}\put(1201,-1561){\circle*{72}}
}%
{\color[rgb]{0,0,0}\put(1201,-1111){\circle*{72}}
}%
{\color[rgb]{0,0,0}\put(1201,-1561){\framebox(900,900){}}
}%
\thicklines
{\color[rgb]{0,0,0}\multiput(1351,-1411)(6.00000,6.00000){26}{\makebox(6.6667,10.0000){\SetFigFont{7}{8.4}{\rmdefault}{\mddefault}{\updefault}.}}
}%
{\color[rgb]{0,0,0}\multiput(1351,-1261)(6.00000,-6.00000){26}{\makebox(6.6667,10.0000){\SetFigFont{7}{8.4}{\rmdefault}{\mddefault}{\updefault}.}}
}%
{\color[rgb]{0,0,0}\multiput(1801,-1411)(6.00000,6.00000){26}{\makebox(6.6667,10.0000){\SetFigFont{7}{8.4}{\rmdefault}{\mddefault}{\updefault}.}}
}%
{\color[rgb]{0,0,0}\multiput(1801,-1261)(6.00000,-6.00000){26}{\makebox(6.6667,10.0000){\SetFigFont{7}{8.4}{\rmdefault}{\mddefault}{\updefault}.}}
}%
{\color[rgb]{0,0,0}\multiput(1801,-811)(6.00000,-6.00000){26}{\makebox(6.6667,10.0000){\SetFigFont{7}{8.4}{\rmdefault}{\mddefault}{\updefault}.}}
}%
{\color[rgb]{0,0,0}\multiput(1801,-961)(6.00000,6.00000){26}{\makebox(6.6667,10.0000){\SetFigFont{7}{8.4}{\rmdefault}{\mddefault}{\updefault}.}}
}%
{\color[rgb]{0,0,0}\multiput(1351,-811)(6.00000,-6.00000){26}{\makebox(6.6667,10.0000){\SetFigFont{7}{8.4}{\rmdefault}{\mddefault}{\updefault}.}}
}%
{\color[rgb]{0,0,0}\multiput(1351,-961)(6.00000,6.00000){26}{\makebox(6.6667,10.0000){\SetFigFont{7}{8.4}{\rmdefault}{\mddefault}{\updefault}.}}
}%
\end{picture}%

%% file: fixpts_20_31
\setlength{\unitlength}{3947sp}%
\begingroup\makeatletter\ifx\SetFigFont\undefined%
\gdef\SetFigFont#1#2#3#4#5{%
  \reset@font\fontsize{#1}{#2pt}%
  \fontfamily{#3}\fontseries{#4}\fontshape{#5}%
  \selectfont}%
\fi\endgroup%
\begin{picture}(1009,1009)(1104,-1658)
\thinlines
{\color[rgb]{0,0,0}\put(1201,-1561){\framebox(900,900){}}
}%
{\color[rgb]{0,0,0}\put(1801,-961){\framebox(150,150){}}
}%
{\color[rgb]{0,0,0}\put(1351,-1411){\framebox(150,150){}}
}%
{\color[rgb]{0,0,0}\put(1801,-1411){\framebox(150,150){}}
}%
{\color[rgb]{0,0,0}\put(1351,-961){\framebox(150,150){}}
}%
\thicklines
{\color[rgb]{0,0,0}\multiput(1126,-1186)(6.00000,6.00000){26}{\makebox(6.6667,10.0000){\SetFigFont{7}{8.4}{\rmdefault}{\mddefault}{\updefault}.}}
}%
{\color[rgb]{0,0,0}\multiput(1126,-1636)(6.00000,6.00000){26}{\makebox(6.6667,10.0000){\SetFigFont{7}{8.4}{\rmdefault}{\mddefault}{\updefault}.}}
}%
{\color[rgb]{0,0,0}\multiput(1576,-1636)(6.00000,6.00000){26}{\makebox(6.6667,10.0000){\SetFigFont{7}{8.4}{\rmdefault}{\mddefault}{\updefault}.}}
}%
{\color[rgb]{0,0,0}\multiput(1576,-1186)(6.00000,6.00000){26}{\makebox(6.6667,10.0000){\SetFigFont{7}{8.4}{\rmdefault}{\mddefault}{\updefault}.}}
}%
{\color[rgb]{0,0,0}\multiput(1576,-1036)(6.00000,-6.00000){26}{\makebox(6.6667,10.0000){\SetFigFont{7}{8.4}{\rmdefault}{\mddefault}{\updefault}.}}
}%
{\color[rgb]{0,0,0}\multiput(1126,-1036)(6.00000,-6.00000){26}{\makebox(6.6667,10.0000){\SetFigFont{7}{8.4}{\rmdefault}{\mddefault}{\updefault}.}}
}%
{\color[rgb]{0,0,0}\multiput(1126,-1486)(6.00000,-6.00000){26}{\makebox(6.6667,10.0000){\SetFigFont{7}{8.4}{\rmdefault}{\mddefault}{\updefault}.}}
}%
{\color[rgb]{0,0,0}\multiput(1576,-1486)(6.00000,-6.00000){26}{\makebox(6.6667,10.0000){\SetFigFont{7}{8.4}{\rmdefault}{\mddefault}{\updefault}.}}
}%
\end{picture}%

%% file: fixpts_10_30
\setlength{\unitlength}{3947sp}%
\begingroup\makeatletter\ifx\SetFigFont\undefined%
\gdef\SetFigFont#1#2#3#4#5{%
  \reset@font\fontsize{#1}{#2pt}%
  \fontfamily{#3}\fontseries{#4}\fontshape{#5}%
  \selectfont}%
\fi\endgroup%
\begin{picture}(1009,1009)(1104,-1658)
{\color[rgb]{0,0,0}\thinlines
\put(1201,-1111){\circle*{72}}
}%
{\color[rgb]{0,0,0}\put(1651,-1111){\circle*{72}}
}%
{\color[rgb]{0,0,0}\put(1201,-1561){\circle*{72}}
}%
{\color[rgb]{0,0,0}\put(1651,-1561){\circle*{72}}
}%
{\color[rgb]{0,0,0}\put(1201,-1561){\framebox(900,900){}}
}%
{\color[rgb]{0,0,0}\put(1126,-1636){\framebox(150,150){}}
}%
{\color[rgb]{0,0,0}\put(1576,-1636){\framebox(150,150){}}
}%
{\color[rgb]{0,0,0}\put(1576,-1186){\framebox(150,150){}}
}%
{\color[rgb]{0,0,0}\put(1126,-1186){\framebox(150,150){}}
}%
\end{picture}%

%% file: fixpts_10_20
\setlength{\unitlength}{3947sp}%
\begingroup\makeatletter\ifx\SetFigFont\undefined%
\gdef\SetFigFont#1#2#3#4#5{%
  \reset@font\fontsize{#1}{#2pt}%
  \fontfamily{#3}\fontseries{#4}\fontshape{#5}%
  \selectfont}%
\fi\endgroup%
\begin{picture}(1009,1009)(1104,-1658)
{\color[rgb]{0,0,0}\thinlines
\put(1651,-1111){\circle*{72}}
}%
{\color[rgb]{0,0,0}\put(1651,-1561){\circle*{72}}
}%
{\color[rgb]{0,0,0}\put(1201,-1561){\circle*{72}}
}%
{\color[rgb]{0,0,0}\put(1201,-1111){\circle*{72}}
}%
{\color[rgb]{0,0,0}\put(1201,-1561){\framebox(900,900){}}
}%
\thicklines
{\color[rgb]{0,0,0}\multiput(1126,-1186)(6.00000,6.00000){26}{\makebox(6.6667,10.0000){\SetFigFont{7}{8.4}{\rmdefault}{\mddefault}{\updefault}.}}
}%
{\color[rgb]{0,0,0}\multiput(1126,-1036)(6.00000,-6.00000){26}{\makebox(6.6667,10.0000){\SetFigFont{7}{8.4}{\rmdefault}{\mddefault}{\updefault}.}}
}%
{\color[rgb]{0,0,0}\multiput(1126,-1486)(6.00000,-6.00000){26}{\makebox(6.6667,10.0000){\SetFigFont{7}{8.4}{\rmdefault}{\mddefault}{\updefault}.}}
}%
{\color[rgb]{0,0,0}\multiput(1126,-1636)(6.00000,6.00000){26}{\makebox(6.6667,10.0000){\SetFigFont{7}{8.4}{\rmdefault}{\mddefault}{\updefault}.}}
}%
{\color[rgb]{0,0,0}\multiput(1576,-1186)(6.00000,6.00000){26}{\makebox(6.6667,10.0000){\SetFigFont{7}{8.4}{\rmdefault}{\mddefault}{\updefault}.}}
}%
{\color[rgb]{0,0,0}\multiput(1576,-1036)(6.00000,-6.00000){26}{\makebox(6.6667,10.0000){\SetFigFont{7}{8.4}{\rmdefault}{\mddefault}{\updefault}.}}
}%
{\color[rgb]{0,0,0}\multiput(1576,-1486)(6.00000,-6.00000){26}{\makebox(6.6667,10.0000){\SetFigFont{7}{8.4}{\rmdefault}{\mddefault}{\updefault}.}}
}%
{\color[rgb]{0,0,0}\multiput(1576,-1636)(6.00000,6.00000){26}{\makebox(6.6667,10.0000){\SetFigFont{7}{8.4}{\rmdefault}{\mddefault}{\updefault}.}}
}%
\end{picture}%

%% file: fixpts_20_30
\setlength{\unitlength}{3947sp}%
\begingroup\makeatletter\ifx\SetFigFont\undefined%
\gdef\SetFigFont#1#2#3#4#5{%
  \reset@font\fontsize{#1}{#2pt}%
  \fontfamily{#3}\fontseries{#4}\fontshape{#5}%
  \selectfont}%
\fi\endgroup%
\begin{picture}(1009,1009)(1104,-1658)
\thinlines
{\color[rgb]{0,0,0}\put(1201,-1561){\framebox(900,900){}}
}%
\thicklines
{\color[rgb]{0,0,0}\multiput(1126,-1186)(6.00000,6.00000){26}{\makebox(6.6667,10.0000){\SetFigFont{7}{8.4}{\rmdefault}{\mddefault}{\updefault}.}}
}%
{\color[rgb]{0,0,0}\multiput(1126,-1636)(6.00000,6.00000){26}{\makebox(6.6667,10.0000){\SetFigFont{7}{8.4}{\rmdefault}{\mddefault}{\updefault}.}}
}%
{\color[rgb]{0,0,0}\multiput(1576,-1636)(6.00000,6.00000){26}{\makebox(6.6667,10.0000){\SetFigFont{7}{8.4}{\rmdefault}{\mddefault}{\updefault}.}}
}%
{\color[rgb]{0,0,0}\multiput(1576,-1186)(6.00000,6.00000){26}{\makebox(6.6667,10.0000){\SetFigFont{7}{8.4}{\rmdefault}{\mddefault}{\updefault}.}}
}%
{\color[rgb]{0,0,0}\multiput(1576,-1036)(6.00000,-6.00000){26}{\makebox(6.6667,10.0000){\SetFigFont{7}{8.4}{\rmdefault}{\mddefault}{\updefault}.}}
}%
{\color[rgb]{0,0,0}\multiput(1126,-1036)(6.00000,-6.00000){26}{\makebox(6.6667,10.0000){\SetFigFont{7}{8.4}{\rmdefault}{\mddefault}{\updefault}.}}
}%
{\color[rgb]{0,0,0}\multiput(1126,-1486)(6.00000,-6.00000){26}{\makebox(6.6667,10.0000){\SetFigFont{7}{8.4}{\rmdefault}{\mddefault}{\updefault}.}}
}%
{\color[rgb]{0,0,0}\multiput(1576,-1486)(6.00000,-6.00000){26}{\makebox(6.6667,10.0000){\SetFigFont{7}{8.4}{\rmdefault}{\mddefault}{\updefault}.}}
}%
\thinlines
{\color[rgb]{0,0,0}\put(1126,-1186){\framebox(150,150){}}
}%
{\color[rgb]{0,0,0}\put(1576,-1186){\framebox(150,150){}}
}%
{\color[rgb]{0,0,0}\put(1126,-1636){\framebox(150,150){}}
}%
{\color[rgb]{0,0,0}\put(1576,-1636){\framebox(150,150){}}
}%
\end{picture}%